\documentclass[11pt]{article}

\usepackage{amsmath}
\usepackage{amssymb}
\usepackage{amsfonts}
\usepackage{mathrsfs}
\usepackage{amsthm}
\usepackage{stmaryrd}
\usepackage{physics}
\usepackage[dvipdfmx]{hyperref}
\usepackage[dvips]{color}
\usepackage{bm}
\usepackage{framed}

\usepackage[top=25truemm,bottom=25truemm,left=20truemm,right=20truemm]{geometry}

\newcommand{\beq}{\begin{equation}}
\newcommand{\eeq}{\end{equation}}
\newcommand{\beqn}{\begin{equation*}}
\newcommand{\eeqn}{\end{equation*}}
\newcommand\al[1]{\begin{align}#1\end{align}}

\newcommand\als[1]{\begin{align}\begin{split}#1\end{split}\end{align}}
\newcommand{\1}{\mbox{1}\hspace{-0.25em}\mbox{l}}

\newcommand {\btheo}{\begin{theo}}
\newcommand {\blemm}{\begin{lemm}}
\newcommand {\bprop}{\begin{prop}}
\newcommand {\bex}{\begin{ex}}
\newcommand {\bprf}{\begin{prf}}
\newcommand {\etheo}{\end{theo}}
\newcommand {\elemm}{\end{lemm}}
\newcommand {\eprop}{\end{prop}}
\newcommand {\eex}{\end{ex}}
\newcommand {\eprf}{\end{prf}}

\renewcommand\thefootnote{\arabic{footnote})}

\makeatletter
  \def\@thmcountersep{-}
\makeatother

\theoremstyle{definition}
\newtheorem{theo}{Theorem}[section]
\newtheorem{lemm}[theo]{Lemma}
\newtheorem{prop}[theo]{Proposition}
\newtheorem{ex}[theo]{Example}
\newtheorem{prf}{Proof}

\def\bea#1\ena{\begin{align}#1\end{align}}
\def\bean#1\enan{\begin{align*}#1\end{align*}} 
\def\p{\partial}

\def\matt[#1,#2,#3,#4]{\left(%
\begin{array}{cc} #1 & #2 \\ #3 & #4 \end{array} \right)}
\def\vect[#1,#2]{\left(%
\begin{array}{cc} #1 \\ #2  \end{array} \right)}
\def\tvect[#1,#2]{\left(%
\begin{array}{cc} #1 & #2  \end{array} \right)}
\def\ket[#1]{| #1 \rangle}
\def\bra[#1]{\langle #1 |}
\def\nn{\nonumber\\}

\def\cM{\mathcal{M}}

\def\slashchar#1{\setbox0=\hbox{$#1$}
\dimen0=\wd0
\setbox1=\hbox{/} \dimen1=\wd1
\ifdim\dimen0>\dimen1
\rlap{\hbox to \dimen0{\hfil/\hfil}}
#1
\else
\rlap{\hbox to \dimen1{\hfil$#1$\hfil}}
/
\fi}
\makeatletter
    
    \@addtoreset{equation}{section}
  \makeatother

\begin{document}

\begin{titlepage}
\renewcommand{\thefootnote}{\fnsymbol{footnote}}
\begin{normalsize}
\begin{flushright}
\begin{tabular}{l}
UTHEP-716
\end{tabular}
\end{flushright}
  \end{normalsize}

~~\\

\vspace*{0cm}
    \begin{Large}
       \begin{center}
         {Information metric, Berry connection and \\
Berezin-Toeplitz quantization for matrix geometry}
       \end{center}
    \end{Large}
\vspace{0.7cm}

\begin{center}
Goro I{\sc shiki}$^{1),2)}$\footnote
            {
e-mail address : 
ishiki@het.ph.tsukuba.ac.jp},
Takaki M{\sc atsumoto}$^{2)}$\footnote
            {
e-mail address : 
matsumoto@het.ph.tsukuba.ac.jp} and 
Hisayoshi M{\sc uraki}$^{3)}$\footnote
            {
e-mail address : 
hmuraki@sogang.ac.kr}

\vspace{0.7cm}

     $^{ 1)}$ {\it Tomonaga Center for the History of the Universe, University of Tsukuba, }\\
               {\it Tsukuba, Ibaraki 305-8571, Japan}\\
                   
     $^{ 2)}$ {\it Graduate School of Pure and Applied Sciences, University of Tsukuba, }\\
               {\it Tsukuba, Ibaraki 305-8571, Japan}\\
     $^{ 3)}$ {\it Department of Physics, Sogang University, }
               {\it Seoul 04107, Korea}\\

\end{center}

\vspace{0.7cm}

\begin{abstract}
\noindent
We consider the information metric and Berry connection
in the context of noncommutative matrix geometry. 
We propose that these objects give a new method of characterizing 
the fuzzy geometry of matrices.
We first give formal definitions of these geometric objects and 
then explicitly calculate them for the well-known matrix 
configurations of fuzzy $S^2$ and fuzzy $S^4$.
We find that the information metrics are given by the usual round metrics for 
both examples, while the Berry connections coincide with
the configurations of the Wu-Yang monopole and the Yang monopole for 
fuzzy $S^2$ and fuzzy $S^4$, respectively.
Then, we demonstrate that 
the matrix configurations of fuzzy $S^n$ $(n=2,4)$ 
can be understood as images of the embedding functions $S^n\rightarrow \textbf{R}^{n+1}$ under the Berezin-Toeplitz quantization map.
Based on this result, we also obtain a mapping rule for the 
Laplacian on fuzzy $S^4$.

\end{abstract}
\vfill

\end{titlepage}
\vfil\eject


\tableofcontents

\section{Introduction}
In the matrix models for string and M- theories
\cite{Banks:1996vh,Ishibashi:1996xs}, geometry of fundamental objects 
such as strings and membranes are described in terms of some Hermitian matrices
$X^\mu$, which correspond to the quantized version of the embedding functions. 
The quantization process to obtain the matrices is very similar to the canonical quantization of classical mechanical systems, 
in which coordinates and conjugate momenta are promoted to 
noncommutative operators acting on a Hilbert space.
In the case of the matrix models, the noncommutativity is introduced 
purely between coordinates and this leads to the notion of 
the noncommutative matrix geometry.

A nice framework of this quantization process is given by the 
matrix regularization \cite{deWit:1988wri,Arnlind:2010ac}.
The matrix regularization can be defined for 
any compact symplectic manifold $({\cal M}, \omega)$  
and is characterized by a sequence $\{T_N \}$, where $N$ runs
over an infinite set of strictly increasing positive integers
and $T_N$ is a linear map from functions on ${\cal M}$ to $N\times N$ 
matrices. Basically, $T_N$ is required to satisfy 
\begin{align}
&||T_N(f)T_N(g)-T_N(f g)|| \rightarrow 0, \nonumber\\
&|| ic_N[T_N(f), T_N(g)]-T_N(\{f, g\})|| \rightarrow 0, 
\nonumber\\
&{\rm Tr}T_N(f) \rightarrow \int \omega^{2n} f,
\label{TN requirements}
\end{align}
 as $N \rightarrow \infty$. Here, $|| \cdot|| $ is a matrix norm, $\{ \; , \; \}$ is a
Poisson bracket on ${\cal M}$ and $c_N$ is an $N$-dependent constant, 
which goes to infinity as $N\rightarrow \infty$ and 
controls the magnitude of noncommutativity.
The first condition in (\ref{TN requirements})
says that the algebra of matrices approximates
the algebra of functions. In particular, it implies that the matrices $T_N(f)$
become commutative in the large-$N$ limit.
The second condition means that, in the large-$N$ limit, the Poisson algebra 
can also be well-approximated by the commutator algebra of matrices.
The third condition for integrals can be used to map 
action functionals on ${\cal M}$ to matrix models.

Though most well-known matrix geometries such as 
the fuzzy $CP^n$, fuzzy tori and so on \cite{Madore:1991bw,Balachandran:2001dd,Connes:1997cr,Arnlind:2006ux} can be regarded as concrete examples of 
the matrix regularization, there are some other examples which 
do not fit into the definition of the matrix regularization. 
In particular, since the definition of the matrix regularization depends on 
the symplectic structure, it can not be applied to 
nonsymplectic manifolds.
For example, $S^4$ is not a symplectic manifold and 
its fuzzy version \cite{Castelino:1997rv} gives a typical 
example which can not be described naively
as the matrix regularization of four-sphere\footnote{
See \cite{Sperling:2017dts} and references therein for various descriptions 
of fuzzy $S^4$.}.
The fact that these nonsymplectic spaces also play
important roles in understanding D-branes in the matrix models 
\cite{Castelino:1997rv} suggests  that 
the requirements of the matrix regularization 
(\ref{TN requirements}) may be too strong,
and more fundamental framework may be necessary to understand fuzzy 
geometries in the matrix models.

In this paper, we consider the Berezin-Toeplitz quantization for 
spinor bundles \cite{Bordemann:1993zv, Ma-Marinescu}. 
This method can be defined on a compact Riemannian 
spin-C manifold equipped with a topologically nontrivial gauge
field configuration. This method provides a systematic 
way of generating a linear map from functions 
on the manifold to $N\times N$ matrices.
Here, the space of $N\times N$ matrices 
arises as a restriction of the functional
space of spinors to the space of Dirac zero modes, where
$N$ is the dimension of the kernel of the Dirac operator and is 
related to the topological charge (such as the monopole charge or 
instanton number) of the gauge field by the index theorem. 
For K\"{a}hler manifolds, this mapping has been shown to satisfy
the properties (\ref{TN requirements}) of the matrix regularization, 
as a consequence of the K\"{a}hler compatibility condition 
\cite{Bordemann:1993zv, Ma-Marinescu}. 
However, since the definition of this quantization depends only 
on the metric and gauge connection, the Berezin-Toeplitz quantization 
map can be defined for nonsymplectic manifolds as well.
Therefore, the Berezin-Toeplitz quantization may serve as 
a more fundamental mathematical framework for matrix models.
Though a lot of concrete matrix configurations
corresponding to various objects in string/M theories 
have been explicitly constructed so far
\cite{Madore:1991bw,Balachandran:2001dd,Connes:1997cr,Arnlind:2006ux}, 
to our best knowledge, 
little work has been done to clarify 
the connection between those configurations and the Berezin-Toeplitz 
quantization scheme\footnote{This quantization 
has been studied in terms of the lowest Landau level problem 
on some monopole backgrounds. See 
\cite{Hasebe:2010vp,Hasebe:2015htd,Hasebe:2017myo}.}.
In this paper, we try to understand this connection more deeply 
to see whether this quantization method indeed gives a good framework 
for matrix models or not.

The problem we consider in this paper is an inverse problem of the 
construction of the Berezin-Toeplitz quantization. 
In the Berezin-Toeplitz quantization, matrices (Toeplitz operators)
are obtained from continuous geometric data, such as manifolds 
and Dirac zero modes\footnote{More precisely, we mean by the geometric data 
the triplet of the manifold, the metric and the gauge connection. 
Note that the Dirac zero modes can be constructed from them.}. 
Conversely, in this paper, we try to 
extract the geometric data
from a given set of Hermitian matrices $X^\mu$ which define a fuzzy space. 
This problem should be particularly important in 
studying the matrix models, which are formulated completely 
in the language of matrices. 

The most important geometric data in the Berezin-Toeplitz quantization 
are the metric and the gauge connection.
In order to recover these geometric objects from the matrices $X^\mu$,
we propose the use of the information metric and Berry connection. 
These objects are calculable from the matrices $X^\mu$ 
and also serve as new objects characterizing the geometry of the fuzzy spaces. 

The definitions of the information metric and Berry connection 
are based on the notion of the coherent states in fuzzy spaces, which 
has been studied in various contexts recently. 
In \cite{Ishiki:2015saa}, the coherent states were introduced 
for fuzzy spaces based on the viewpoint that 
they should have minimal wave packets in the target space
in the large-$N$ limit. 
In this formulation, the coherent states are 
defined as ground states of a certain Hamiltonian.
This construction was then generalized to the case of finite-$N$
matrices \cite{Schneiderbauer:2016wub}. 
In the earlier work \cite{Berenstein:2012ts},
the use of Dirac operator on 
D0-branes was proposed based on a string-theory viewpoint. 
This method also leads to the notion of the coherent states. 
Since in all of these formulations, coherent states play very important 
roles, we call these methods collectively the coherent state methods
in this paper.
See \cite{deBadyn:2015sca,Karczmarek:2015gda,Ishiki:2016yjp}
 for some analysis using the coherent state methods.
See also \cite{TachyonA, TachyonB} for a nice interpretation of 
the coherent state methods in the system of non-BPS D-branes.

Based on the coherent state methods, one can 
define the information metric and Berry connection for fuzzy spaces. 
In this paper, we first give 
formal definitions of these geometric objects.
Then, we calculate the objects explicitly
for fuzzy $S^2$ and fuzzy$S^4$ as examples.
Finally, we show that for both cases the coherent states form a basis 
of the Dirac zero modes, so that the Hilbert space on which the matrices are 
acting can be identified with the space of the Dirac zero modes in the 
Berezin-Toeplitz quantization.
We also demonstrate that, under the Berezin-Toeplitz quantization map,
the defining Hermitian matrices for
the fuzzy $S^2$ and $S^4$ can be seen as the images 
of the embedding functions of $S^2$ and $S^4$, respectively, 
into the flat target spaces.
This result provides a unified viewpoint for fuzzy $S^2$ and fuzzy $S^4$.
By using the quantization map, we also obtain explicit mapping 
rules for Laplacians on $S^2$ and $S^4$.

This paper is organized as follows.
In section~\ref{Coherent states and classical geometries}, 
we review the coherent state methods.
In section~\ref{Information metric and Berry connection}, 
we introduce the information metric and Berry connection. 
In section~\ref{Examples}, 
we compute these structures for fuzzy $S^2$ and $S^4$.
In section~\ref{Berezin-Toeplitz quantization}, 
we first review the Berezin-Toeplitz quantization and
then show that the matrix configurations of fuzzy $S^2$ and 
fuzzy $S^4$ can be interpreted as the images of 
the embedding functions on $S^2$ and $S^4$, respectively.
In section~\ref{Summary and discussion}, 
we summarize our results. 

\section{Coherent state methods}
\label{Coherent states and classical geometries}
For a given set of Hermitian matrices $\{X^\mu\}$,
we can define an analogue of the coherent states. By using them, we 
can then associate the corresponding commutative space ${\cal M}$ with 
the given matrices. 
In this section, we review this process for two different methods 
based on the Hamiltonian and the Dirac operator.


\subsection{Hamiltonian}


We start with a set of $N\times N$ 
Hermitian matrices $\{X^\mu\} \ (\mu=1,\cdots,D)$, 
which defines a fuzzy space. We assume that there exists 
a commutative limit such that $X^\mu$ become mutually commuting and 
this limit is given by the large-$N$ limit\footnote{
For example, for the fuzzy sphere, 
$D=3$ and $X^\mu $ are given by the $N$-dimensional irreducible 
representation matrices $L_i (i=1,2,3) $ of the $SU(2)$ generators.
For a unit sphere, $X^i$ should be normalized to satisfy $(X^i)^2 = \1_N $ and 
thus $X^i = \frac{2}{\sqrt{N^2-1}}L_i$. With this normalization, 
$X^i$ become commuting matrices in the large-$N$ limit.}.
We also call this limit the classical limit by analogy with 
the quantum mechanics, where the commutative limit $\hbar \rightarrow 0$ 
indeed corresponds to the classical limit.
In terms of the matrix regularization, this setup corresponds to a 
situation such that we are first given the images 
$X^\mu := T_N(x^\mu)$ of the embedding functions 
$x^\mu : {\cal M} \rightarrow \textbf{R}^D$.
But the following arguments apply not only to the symplectic manifolds 
but also to nonsymplectic manifolds such as $S^4$. In 
the latter case, one can also construct the corresponding matrices 
$X^\mu$ based on the observations of D-brane charges or symmetries
\cite{Castelino:1997rv}.

For the given $N\times N$ Hermitian matrices $\{X^\mu\}$, 
we first introduce the Hamiltonian, which is an $N\times N$ Hermitian 
matrix defined by
\beq
	\label{1.2}
	H(y)=\frac{1}{2}\sum_{\mu=1}^D(X_{\mu}-y_{\mu}\1_{N})^2.
\eeq
Here $y_{\mu}\,(\mu=1,2,\cdots,D)$ are real parameters and 
$\1_{N}$ stands for the $N\times N$ identity matrix\footnote{In 
the following, we omit the $N\times N$ identity matrix $\1_N$ 
for notational simplicity.}.
Since the Hamiltonian $H(y)$ is Hermitian for any $y$,
it is always possible to diagonalize $H(y)$ by using 
unitary similarity transformations.
We introduce a basis, on which $H(y)$ becomes diagonal:
\beq
	\label{eigenHam}
		H(y)|n,y\rangle=E_n(y)|n,y\rangle,\ \ (n=0,1,\cdots,N-1).
\eeq
Since $H(y)$ is a non-negative matrix, all the eigenvalues $E_n(y)$ are
non-negative. We label the eigenvalues as
$ 0\leq E_0(y) \leq E_1(y) \leq\cdots\leq E_{N-1}(y)$.
The eigenstates shall be normalized as $\langle n,y|m,y\rangle=\delta_{nm}$.

In quantum mechanics, the canonical coherent states are the states with 
minimal wave packets and in particular, the sizes of the wave packets 
go to zero in the classical limit $\hbar \rightarrow 0$. 
For fuzzy spaces defined by $\{X^\mu\}$, we can 
introduce an analogue of the canonical coherent states by using the 
above Hamiltonian as follows. 
From the definition of the eigenstates, 
the lowest eigenvalue $E_0(y)$ can be expressed as
\beq
	\label{1.3}
	E_0(y)=\bra[0,y]H(y)\ket[0,y]
	=\frac{1}{2}(\langle X_{\mu}\rangle-y_{\mu})^2+\frac{1}{2}(\Delta X_{\mu})^2,
\eeq
where 
\bea
	&\langle X_{\mu}\rangle = \bra[0,y]X_\mu\ket[0,y],\\
	&\Delta X_{\mu} = \sqrt{\bra[0,y]X^2_\mu\ket[0,y]-\bra[0,y]X_\mu\ket[0,y]^2}.
\ena
In terms of the wave packet of the ground state $|0, y \rangle$, 
$\langle X_{\mu}\rangle $ corresponds to 
the position of the center of the wave packet in the target space, 
while $\Delta X_{\mu}$ corresponds to the size of the wave packet 
in the $\mu$ direction.
Now, suppose that $E_0(y)$ for a certain $y$ goes to zero in the 
classical (commutative) limit as $E_0(y) \rightarrow 0$.
Then, since both terms in the right-hand side of \eqref{1.3} are
squared and positive, we have
\bea
	\begin{cases}
		\langle X_{\mu}\rangle-y_{\mu}  \longrightarrow  0 ,\\
		\Delta X_{\mu}  \longrightarrow  0.
	\end{cases}
	\label{conds}
\ena
for all $\mu$ simultaneously.
This means that at the point $y$, there exists a wave packet which can shrink 
to zero size in the classical limit. 
Note that the inverse statement is also true, namely, if there is a state 
which (is not necessarily an eigenstate of $H(y)$ but) satisfies 
(\ref{conds}) for a certain point $y \in \textbf{R}^D$, 
the ground state energy $E_0(y)$ is vanishing in the classical limit. 
Thus, the zero loci of $E_0(y)$ in $\textbf{R}^D$ is equivalent to the subspace of $\textbf{R}^D$
such that there can exist a wave packet which shrinks to a point in the 
classical limit. 
Such states are counter objects of the canonical coherent states in quantum 
mechanics.  
From this analogy, we call $|0,y \rangle $ coherent states here 
if it satisfies (\ref{conds}). 

For the fuzzy space defined by $\{X^\mu \}$, 
we can associate a classical (commutative) manifold $\cM$ as 
a hypersurface in $\textbf{R}^D$ defined as a set of points on which 
there exist coherent states.
In other words, $\cM$ is given by the zero loci of the Hamiltonian:
\beq
	\label{1.4}
	\mathcal{M}
	=\left\{y\in\textbf{R}^D\,|\,f(y)=0\right\},
\eeq
where we introduced a function 
\bea
	f(y)={\lim_{N\to\infty}}E_0(y).
\ena

Note that, in most cases of finite-size matrices, exact zero modes of the 
Hamiltonian do not exist, and the classical space ${\cal M}$ can only be 
defined with the large-$N$ limit in this method.
However, this method can be extended to finite-$N$ cases with 
the use of quasi coherent states \cite{Schneiderbauer:2016wub}.
Note also that even in the large-$N$ limit, 
if we consider general matrices for $X^\mu$, there are a lot of cases 
where ${\cal M}$ becomes an empty set. In order to have a non empty set,
$\{X^\mu \}$ need to become commutative in the large-$N$ limit as 
$[X^\mu, X^\nu] \rightarrow 0$. 

In summary, we first introduced the coherent states as 
the ground state eigenvectors of the Hamiltonian (\ref{1.2})
which have vanishing eigenvalues in the large-$N$ limit.
Next, we defined a classical space ${\cal M}$ as a set of points in $\textbf{R}^D$
on which there exists the coherent states. 

\subsection{Dirac Operator}

We next introduce another method based on a matrix Dirac-type operator
\cite{Berenstein:2012ts, TachyonA, TachyonB}.
While the method using Hamiltonian is based on the analogy with the quantum 
mechanics, the method using Dirac operator is based on some observations in 
string theories. Here, we first show the mathematical treatment of this 
method and then explain its physical implications in string theory.

The Dirac operator is defined from the given matrix $X^\mu$ as
\bea  
	\slashchar{D}(y)
	=\delta_{\mu\nu}{\Gamma^\mu}\otimes({X^\nu}-{y^\nu}\1_{N}).
\label{Def Dirac}
\ena
This is a $2^{[D/2]}N\times2^{[D/2]}N$ 
Hermitian matrix, where $[D/2]$ 
stands for the maximal integer less than or equal to $D/2$.
Here $\Gamma^\mu$ are $2^{[D/2]}\times2^{[D/2]}$ matrix representations of the 
$D$-dimensional Euclidean Clifford algebra:
\bea
	\{\Gamma^\mu,\Gamma^\nu\}=2\delta^{\mu\nu}\1_{2^{[D/2]}}.
\ena

The classical space ${\cal M}$ is defined as a hypersurface 
on which there exist zero modes of the Dirac operator as follows.
Since the Dirac operator is Hermitian matrix, it has real eigenvalues. 
We denote eigenvalues and eigenstates as
\beq
	\label{eigenDirac}
		\slashchar{D}(y)|n,y\rangle=E_n(y)|n,y\rangle,\ \ (n=0,1,\cdots,2^{[D/2]}N-1),
\eeq
where we order the eigenvalues as 
$|E_0(y)|\leq|E_1(y)|\leq\cdots\leq|E_{2^{[D/2]}N-1}(y)|$ and the 
eigenstates shall be normalized as $\langle n,y|m,y\rangle=\delta_{nm}$.
Note that $E_n(y)$ can also take negative values unlike the case of the 
Hamiltonian. 
The classical space ${\cal M}$ is defined as hypersurfaces on 
which zero modes of the Dirac operator exist:
\bea
	\mathcal{M}=\{y \in\textbf{R}^D\,|\, E_0(y)=0\} .\label{ClHyp}
\ena
This definition of the classical space may look similar to that using the Hamiltonian \eqref{1.4}.
A crucial difference is that in the method using the Dirac operator, 
we do not need to take the large-$N$ limit to define the classical space 
${\cal M}$. The Dirac operator allows exact zero modes even for finite $N$, and 
the geometry is defined for a fixed finite $N$.

There are two different interpretations of this construction in the context of
the string theory. One is based on the probe picture of D0-brane action
\cite{Berenstein:2012ts}. 
Suppose $N$ D0-branes form a bound state such as fuzzy sphere and behave 
as higher dimensional D-brane. Let $X^\mu$ be the matrix configuration 
(the bosonic fields) of these D0-branes. In addition, 
we consider another probe D0-brane at $y^\mu$. 
Then, the Dirac operator (\ref{Def Dirac})
appears in the fermionic
kinetic term of the open string modes connecting the bounded 
D0-branes and the probe brane. $E_0$ measures the 
lowest energy of the open string, which is in general proportional to the 
length of the open string. Thus, at the position where 
Dirac zero modes exist, the probe brane hits the D0-branes, and 
hence 
${\cal M}$ defined by (\ref{ClHyp}) gives the geometry of the 
D0-branes seen by the probe brane.

The second interpretation is provided by flat non-BPS D-brane systems
in superstring theory \cite{TachyonA, TachyonB} (see also 
\cite{Terashima:2005ic,Asakawa:2001vm}). 
The theory on the non-BPS D-branes generally contains 
the tachyon field $T(y)$, and the potential term of $T(y)$ in 
the low energy action is proportional to 
the exponential factor $e^{-T(y)^2}$.
The theory possesses a classical solution, which represents 
tachyon condensations. The solution takes the form 
$T(y)=u \slashchar{D}(y)$, where $X^\mu$ in (\ref{Def Dirac}) can 
be arbitrary constant Hermitian 
matrices. In order for this to be a solution of the equation of motion,
the parameter $u$ has to be sent to infinity.
Then, since the potential energy is proportional to 
$e^{-u^2\slashchar{D}(y)^2 }$ with $u \rightarrow \infty$, only 
zero modes of the Dirac operator survive. 
In particular, this is possible only when $y\in {\cal M}$, where 
${\cal M}$ is defined by (\ref{ClHyp}). 
Thus, this solution corresponds to a situation that 
the original non-BPS branes with the world volume 
coordinates $y^\mu$ becomes another configurations of D-branes which 
has the shape of ${\cal M}$. From the analysis of the boundary 
string field theory, the latter D-branes are found to be stable 
BPS D-branes. Thus, in this context, ${\cal M}$ given by 
(\ref{ClHyp}) corresponds
to the shape of the BPS D-branes produced after the tachyon condensation.


%
The square of  the Dirac operator is calculated as
\bea
	\label{SqOfDir}
	\slashchar{D}^2(y)
	&=\1_{2^{[D/2]}}\otimes({X^\mu}-{y^\mu}I_{N})^2
		+\frac{1}{4}[\Gamma^\mu,\Gamma^\nu]\otimes[{X^\mu},{X^\nu}].
\ena
Note that the first term on the right-hand side is proportional to 
the Hamiltonian. For the commuting matrices in the large-$N$ limit, 
we have
\bea
	\slashchar{D}^2(y)
	& \simeq 2\1_{2^{[D/2]}}\otimes H(y)
	\label{DirSq}
\ena
in the large-$N$ limit.
Thus, the Dirac operator asymptotically coincides with
the Hamiltonian in the large-$N$ limit. One may think that the 
relation (\ref{DirSq}) 
between the Hamiltonian and the Dirac operator in the large-$N$ limit
also implies the equivalence of the classical spaces defined by the two 
methods. 
However, rigorously speaking, there are some cases in which 
the classical spaces do not coincide with each other.
Let us denote by ${\cal M}_H$ and ${\cal M}_{\slashchar{D}}$ the 
classical spaces defined by the Hamiltonian and the Dirac operator 
in the large-$N$ limit, respectively. 
For $y^\mu \in {\cal M}_{\slashchar{D}}$, there exists 
a zero mode of the Dirac operator as $\slashchar{D}(y)|0,y\rangle=0$.
Then we have
\bea
	0      &=\bra[0,y] \slashchar{D}^2(y)|0,y\rangle 
\simeq 2 \bra[0,y] \1_{2^{[D/2]}}\otimes H(y)|0,y\rangle.
\ena
Thus, there also exists a zero mode of the Hamiltonian at the 
same point $y^\mu$ in the large-$N$ limit. This means that 
${\cal M}_{\slashchar{D}} \subset {\cal M}_H$.
However, the inverse statement may not be true in general. 
For $y^\mu \in {\cal M}_H$, one can only say that 
$\slashchar{D}$ has an approximate zero mode in the large-$N$ limit,
namely, $\slashchar{D}$ may not have an exact zero mode at that point.
Thus, if there exists a point at which any eigenvalues of $\slashchar{D}$
are nonzero but some of them are very small as ${\cal O}(1/N)$, 
such point will be included in ${\cal M}_H$ but not
in ${\cal M}_{\slashchar{D}}$. Hence, in general, we have only
the relation ${\cal M}_{\slashchar{D}} \subset {\cal M}_H$.

The method using the Dirac operator has an advantage that the geometry 
can be rigidly defined at finite $N$ and hence mathematically 
rigorous treatment is possible at finite $N$. 
However, depending on a context, 
one may be interested in the geometry which emerges only
in the large-$N$ limit. The method using the Hamiltonian 
has an advantage that,
without introducing the vector space of spinors, 
one can pick up not only the points in ${\cal M}_{\slashchar{D}}$
but also approximately emergent points.

In the both pictures, 
each zero eigenstate describes a single D-brane.
If there are some degenerate zero eigenstates of the Dirac operator,
they corresponds to multiple coincident D-branes. 


\section{Information metric and Berry connection}
\label{Information metric and Berry connection}
In this section, we give definitions of the information metric 
and Berry connection on $\mathcal{M}$.
In this section, we use the Dirac operator method, but the same 
arguments apply to the Hamiltonian method as well.

\subsection{Information Metric}
Suppose that the Dirac operator has $k$ degenerate zero modes 
and the zero eigenstates are labeled as $\ket[0,a,y]$ 
$(a=1,2,\cdots,k)$.
From these eigenstates, we first define a density matrix,
\bea
	\rho(y) = \frac{1}{k}\sum_{a=1}^k \ket[0,a,y]\bra[0,a,y],	
\label{densitymatdeg}
\ena
which is proportional to the projection operator onto the 
$k$-dimensional vector space spanned by $\ket[0,a,y]$.
Note that $\rho(y)$ is the unique density matrix made of 
the zero eigenstates and invariant under the 
$U(k)$ rotational transformation of the zero eigenstates, 
\bea
	\ket[0,a,y] \ \mapsto \ \sum_{b=1}^k
\ket[0,b,y] V_{ba}(y), \ \ V(y)\in U(k).
\label{Uk rotation}
\ena

We consider the case that ${\cal M}$ defined in (\ref{ClHyp}) 
is a smooth simply connected compact manifold corresponding to
extended D-branes\footnote{In general, ${\cal M}$ contains 
some disconnected components. 
The following arguments can be easily extended to such general cases.}. 
On a vicinity of this manifold, 
$|0,a,y \rangle$ are differentiable\footnote{
Note that $|0,a,y+\epsilon \rangle = 
|0,a,y \rangle + \epsilon^\mu \partial_\mu |0,a,y \rangle + \cdots$. 
The derivative terms are explicitly given by the formula of the 
perturbation theory under 
$\slashchar{D}(y+\epsilon) =\slashchar{D}(y) - \Gamma^\mu \epsilon_\mu$.
This perturbation should be smooth at least when $\epsilon^\mu$ 
is much smaller than the spectral gap of the Dirac operator.}.
Then, $\rho$ defined by (\ref{densitymatdeg}) gives a smooth map 
from ${\cal M}$ to the space of the density matrices.
Furthermore, we can show that $\rho$ and its differential $d\rho$ are injective
mappings. See appendix \ref{Injectivity of rho and drho} for our proof.
Then, $\rho$ gives an embedding of ${\cal M}$ into the space 
$\mathcal{D}$ of all density matrices,  
\bea
	\rho : \cM \longrightarrow 
	\mathcal{D}.
	\label{EmbdDMat}
\ena

The space $\mathcal{D}$ of the density matrices forms a convex cone 
and one can define a metric structure on this space.
In fact, the information (Bures) metric provides a natural metric 
on $\mathcal{D}$, 
defined by
\begin{align}
ds^2 = \frac{1}{2}{\rm Tr} (d \rho G),   \;\;\;  d \rho = \rho G + G \rho.
\label{inf met}
\end{align}
Here, the trace is taken over the vector space associated with the density 
matrices and $G$ is defined from $\rho$ by the second equation of 
(\ref{inf met}).
For pure states, the information metric is 
equivalent to the Fubini-Study metric on the complex projective space 
given by a set of normalized complex vectors.

The embedding (\ref{EmbdDMat}) then defines the pullback of the 
information metric. This pullback provides a metric structure for ${\cal M}$.
For $\rho$ given by \eqref{densitymatdeg}, differentiating 
the relation $\rho^2 = \rho /k$, one finds that
\beq
	G=kd \rho.
\eeq
The pullback can then be explicitly written as
\bea
ds^2 &=
\frac{1}{k}\left(
\sum_{a=1}^{k}  || d| 0,a,y \rangle ||^2
-\sum_{a,b=1}^{k}  \left| \bra[0,a,y] d |0,b,y \rangle \right|^2
\right).
\label{InformMet}
\ena
Here $|| \cdot ||$ is the vector norm and $d \ket[0,a,y ]$ is understood as
\bea
	\ket[0,a,dy ] =  \frac{\partial }{\partial \sigma^\alpha} 
\ket[0,a,y ] d\sigma^\alpha ,
\label{def of dket}
\ena
where $\{ \sigma^\alpha \}$ is arbitrary local coordinate 
on ${\cal M}$. The local coordinate should be chosen such that 
$\{ \sigma^\alpha \}$ parameterizes the zeros 
of $\slashchar{D}(y)$ as $E_0(y(\sigma))=0$.


\subsection{Berry connection}

We can consider a gauge connection on ${\cal M}$ associated 
with the local $U(k)$ rotation (\ref{Uk rotation})
of the zero eigenstates $|0, a, y \rangle$.
This gauge field corresponds to the (non-Abelian) Berry connection. 
The Berry connection is defined as the following one form on ${\cal M}$:
\bea
	\label{GFBeryyUk}
	A_{ab}(\sigma) = -i \bra[0,a,y]d |0,b,y\rangle,
\ena
where $d |0,b,y\rangle$ is defined in (\ref{def of dket}).
It is easy to see that (\ref{GFBeryyUk}) transforms as a non-Abelian 
gauge field under the transformation (\ref{Uk rotation}) as
\bea
A \rightarrow V^\dagger A V -i V^\dagger dV.
\ena

For well-known fuzzy spaces such as fuzzy $S^2$ and $S^4$, this 
gauge field takes topologically nontrivial configurations such as 
monopoles and instantons. We demonstrate this calculation in the 
following sections. 

Let us comment on the setup considered in \cite{Ishiki:2016yjp},
in which matrices $\{ X^\mu \}$ behave as 
\begin{align}
[X^\mu, X^\nu] = \frac{i}{c_N}W^{\mu \nu}(X) + \cdots. 
\label{comxx}
\end{align}
Here, $c_N$ is an $N$-dependent constant which goes to infinity 
in the large-$N$ limit, and 
$\cdots$ represents higher order terms in $1/c_N$.
$W^{\mu \nu}(X)$ in (\ref{comxx}) is antisymmetric in the 
indices $\mu, \nu$ and is 
a polynomial in $X^\mu$ with convergent degree and coefficients in the 
large-$N$ limit.  
For the matrices satisfying (\ref{comxx}), 
It was shown in \cite{Ishiki:2016yjp} 
that the curvature 2-form of the 
Berry connection gives a symplectic form in the large-$N$ limit. 
Namely, the curvature 2-form is closed and non-degenerate.
The information metric was also shown to be the compatible 
K\"{a}hler metric for the symplectic form.

In the setup with D-branes studied in
\cite{Berenstein:2012ts, TachyonA, TachyonB}, 
the Berry connection is understood 
as the gauge field on the D-branes, as first noted in 
\cite{Terashima:2005ic,Hashimoto:2005qh}. 
For K\"{a}hler manifolds, the information metric 
is the compatible world volume metric on the D-branes.

\section{Examples}
\label{Examples}

In this section, 
we consider fuzzy $S^2$ and $S^4$ as examples. 
Through explicit calculations, 
we demonstrate that the information metric for these spaces are 
given by the ordinary round metric, while the 
Berry connections are given by the configurations of 
the Wu-Yang monopole and Yang monopole for fuzzy 
$S^2$ and $S^4$,
respectively.

\subsection{Fuzzy $S^2$}

\subsubsection{Definition of fuzzy $S^2$}

In the standard description of the fuzzy $S^2$, 
one uses three Hermitian matrices, which correspond to 
the quantized embedding functions into $\textbf{R}^3$.
The three matrices are given as 
\bea 
	X^i = R L^i,   \;\;\; R= \frac{2}{\sqrt{N^2-1}}.
\label{X for fuzzy s2}
\ena
Here, $L^i$ are the $SU(2)$ generators in the spin-$J$ 
irreducible representation, where $J$ is related to the matrix size $N$
by $N=2J+1$.
The normalization factor $R$ is chosen so that the fuzzy sphere 
has a unit radius as 
\bea
	\sum_{i=1}^3 (X^i)^2=1.
\ena
These matrices satisfy the commutation relations\footnote{
Note that this commutation relation is of the form of (\ref{comxx}).
Thus, the argument in \cite{Ishiki:2016yjp} can be applied. } 
\bea
	[X^i,X^j]=\frac{2i\epsilon^{kij}}{\sqrt{N^2-1}}X^k.
\ena

For later convenience, we introduce the standard basis 
$|J, m \rangle $ of the 
representation space of $L^i$. They satisfy 
\begin{align}
& L^3 | J,m \rangle = m | J,m \rangle, \nonumber\\
& L^\pm | J,m \rangle = \sqrt{(J\mp m) (J\pm m +1)} | J, m\pm 1 \rangle, 
\label{def of states}
\end{align}
where $L^\pm = L^1 \pm i L^2$.
For $J=1/2$, we also use the shorthand notation,
\begin{align}
|1/2, \pm \rangle := | 1/2, \pm 1/2 \rangle.
\end{align}

\subsubsection{Classical space for fuzzy $S^2$}

The Dirac operator for fuzzy $S^2$ is given by 
a $2N\times2N$ Hermitian matrix,
\begin{align}
	\slashchar{D}(y)
	=\sigma^i\otimes(R L^i-y^i).
\label{Dirac operator for fuzzy S2}
\end{align}
Here $\sigma^i$ are the Pauli matrices. 
The spectrum of (\ref{Dirac operator for fuzzy S2}) 
is derived in the appendix~\ref{ApE2} 
(See also \cite{deBadyn:2015sca,Karczmarek:2015gda,TachyonA}).
There are three types of the eigenstates, 
\begin{align}
	&\ket[\psi^{(\pm)}_m] = (U_2\otimes U_N)
	\big(a^{(\pm)}_m \ket[1/2,+]\otimes \ket[J,m] +b^{(\pm)}_m \ket[1/2,-]\otimes \ket[J,m+1] \big), \nonumber\\
	&\ket[\psi_J] = (U_2\otimes U_N)
	\ket[1/2,+]\otimes \ket[J,J] \ , \nonumber \\
	&\ket[\psi_\bullet] = (U_2\otimes U_N) \ket[1/2,-]\otimes \ket[J,-J],
\label{eigenstates of fuzzy S2}
\end{align}
where $m=-J,-J+1,\cdots, J-2,J-1$, and the corresponding
eigenvalues are given by 
\bea
\lambda^{(\pm)}_m(y) &= - \frac{R}{2} \pm \frac{1}{2}
\sqrt{R^2+4\{|y|^2 -R(2m+1)|y| +R^2J(J + 1)\}}, 
\nonumber\\
\lambda_J(y)&=RJ-|y|,
\nonumber\\
\lambda_\bullet (y)&=RJ+|y|, \label{EignVal}
\ena
respectively, where $|y|= \sqrt{\sum_i y_i^2}$.
In (\ref{eigenstates of fuzzy S2}), 
$a_m^{(\pm)}$, $b_m^{(\pm)}$ are real coefficients satisfying
\bea
	a^{(\pm)}_m= -\frac{R \sqrt{(J-m)(J+m+1)}}{Rm-|y|-\lambda^{(\pm)}_m}b^{(\pm)}_m
\label{equation for am and bm}
\ena
as well as the normalization condition 
\bea
	{a^{(\pm)}}_m^2 + {b^{(\pm)}}_m^2=1 .
	\label{normaliz}
\ena
The unitary matrices $U_2$ and $U_N$ 
in (\ref{eigenstates of fuzzy S2}) are defined in appendix\ref{ApE}.

Note that $\lambda_\bullet (y)$ is strictly positive and 
$\lambda^{(\pm)}_m(y)$ cannot be zero for $m=-J,-J+1,\cdots ,J-1$.
Thus, only $\ket[\psi_J]$ can be the zero mode of the Dirac operator. 
The classical space is then defined as a set of $y \in \textbf{R}^3$ on which 
the zero mode exists:
\bea
	\mathcal{M}=\{y^i\in\textbf{R}^3\,|\, |y|= RJ\} \ .
\label{cl sp for fuzzy sphere}
\ena
Obviously, the classical space is given by a sphere with the radius $RJ$
embedded in $\textbf{R}^3$.
We can also apply the method using the Hamiltonian. 
This is shown in appendix 
\ref{appendix Hamiltonian method s2}.




\subsubsection{Information metric and Berry connection for fuzzy $S^2$}
\label{Information metric and Berry connection for fuzzy S2}
We parametrize the classical space
(\ref{cl sp for fuzzy sphere}) by 
\begin{align}
y^1 &= RJ \sin\theta \cos\phi, \nonumber\\
y^2 &= RJ \sin\theta \sin\phi, \nonumber\\
y^3 &= RJ \cos\theta.
\end{align}
We also introduce the stereographic coordinate $(z,z)$ on 
the classical space by 
\begin{align}
z= e^{i\phi} \tan \frac{\theta}{2}.
\label{stereographic coordinate}
\end{align}

We first compute the information metric (\ref{InformMet}) for 
the zero eigenstate $|\psi_J \rangle$ 
for the fuzzy $S^2$. The expression 
(\ref{form of U 2}) for the unitary matrix $U$ is very useful 
in computing the 
differential of the zero mode $|\psi_J \rangle$. 
In the stereographic coordinate, the differential of 
$U_N (y) |J,J \rangle$ is given by
\begin{align}
d U_N (y) |J,J \rangle = 
\frac{J(\bar{z}dz -z d\bar{z})}{1+|z|^2} U_N(y) |J,J \rangle
+\frac{\sqrt{2J}dz}{1+|z|^2} U_N(y)|J,J-1 \rangle.
\label{dpsi}
\end{align}
By using this, we can easily  compute (\ref{InformMet}). 
The result is given by
\begin{align}
ds^2 = || d| \psi_J \rangle ||^2 - |\langle \psi_J| d|\psi_J\rangle  |^2
=N \frac{dzd\bar{z}}{(1+|z|^2)^2}.
\end{align}
This is nothing but a round K\"{a}hler metric for $S^2$. 
The overall factor 
also picks up information of the density of D0-branes, which is 
an intrinsic data of the matrices $X^\mu$.

By using (\ref{dpsi}), we can also compute the Berry connection 
(\ref{GFBeryyUk}) for fuzzy $S^2$. 
The result is given by
\begin{align}
A = -i \langle \psi_J | d|\psi_J \rangle 
= -i \frac{N}{2}\frac{\bar{z}dz -z d\bar{z}}{1+|z|^2}.
\end{align}
This is just the Dirac monopole configuration. 
The field strength is 
\begin{align}
F= dA
= iN \frac{dz \wedge d\bar{z}}{(1+|z|^2)^2}.
\label{F for fuzzy S2}
\end{align}
The monopole flux (or equivalently the first Chern class) 
coincides with the matrix size $N$:
\begin{align}
\frac{1}{2\pi} \int F = N.
\end{align}

\subsection{Fuzzy $S^4$}

\subsubsection{Definition of fuzzy $S^4$}

Let us first introduce the following orthonormal vectors\footnote{See 
also 
\cite{Castelino:1997rv,Guralnik:2000pb} for the calculation in 
this subsection.}:
\begin{align}
|\eta_1 \rangle =
\left( 
\begin{array}{c}
1 \\
0 \\
0 \\
0 \\
\end{array} 
\right), \;\;\; 
|\eta_2 \rangle =
\left( 
\begin{array}{c}
0 \\
1 \\
0 \\
0 \\
\end{array} 
\right), \;\;\; 
|\eta_3 \rangle =
\left( 
\begin{array}{c}
0 \\
0 \\
1 \\
0 \\
\end{array} 
\right), \;\;\; 
|\eta_4 \rangle =
\left( 
\begin{array}{c}
0 \\
0 \\
0 \\
1 \\
\end{array} 
\right).
\end{align}
We denote by ${\cal H}_n$ the Hilbert space spanned by all 
$n$-fold totally symmetric tensor products of 
$|\eta_i \rangle$ $(i=1,2,3,4)$.
We denote by $N$ the dimension of this space:
\begin{align}
N={\rm dim}{\cal H}_n =
\left(
\begin{array}{c}
n+3 \\
3 \\
\end{array}
\right)
= \frac{(n+1)(n+2)(n+3)}{6}.
\label{dim of hn}
\end{align}
We also denote by ${\cal H}^+_n$ the subspace of ${\cal H}_n$ spanned 
only by all symmetric tensor products of $|\eta_1 \rangle $ and 
$| \eta_2 \rangle$. 
The dimension of this subspace is 
\begin{align}
{\rm dim}{\cal H}^+_n =
\left(
\begin{array}{c}
n+1 \\
1 \\
\end{array}
\right)
= n+1.
\end{align}
We also introduce the five dimensional gamma matrices $\Gamma^A$, 
satisfying $\{\Gamma^A, \Gamma^B \}=2\delta^{AB}\1_4$.
In the following, we use the following representation of $\Gamma_A$:
\als
{
	\Gamma_i&=\sigma_2\otimes\sigma_i
	=\mqty(0&-i\sigma_i \\ i\sigma_i&0)\quad i\in\qty{1,2,3}, \\
	\Gamma_4&=\sigma_1\otimes\1_2=\mqty(0&\1_2 \\ \1_2&0), \\
	\Gamma_5&=\sigma_3\otimes\1_2=\mqty(\1_2&0 \\ 0&-\1_2),
\label{def of gamma}
}
where $\sigma_i$ are the Pauli matrices. 
Note that $|\eta_i \rangle $ are eigenvectors of $\Gamma_5$ with
$ \Gamma_5| \eta_{1,2} \rangle = | \eta_{1,2} \rangle$ and 
 $ \Gamma_5| \eta_{3,4} \rangle = -| \eta_{3,4} \rangle$.

The vector space ${\cal H}_n $ gives the $N$-dimensional 
irreducible representation space of the $SO(5)$ Lie group. 
The $SO(5)$ generators $\Sigma_{AB}$ $(A,B=1,2,\cdots,5)$ 
are represented on ${\cal H}_n$ as
\begin{align}
D_{{\cal H}_n}(\Sigma_{AB}) = \frac{1}{2}(
\Gamma_{AB} \otimes\1_4\otimes\cdots\otimes\1_4
+\1_4 \otimes \Gamma_{AB} \otimes \cdots\otimes\1_4
	+\cdots+\1_4\otimes\cdots\otimes\1_4\otimes \Gamma_{AB}).
\label{symmetrized generator}
\end{align}

The fuzzy $S^4$ is defined by the configuration of the five matrices on 
${\cal H}_n$,
\beq
	\label{Matrix for fuzzy S4}
	X_A=R G^{(n)}_A, \quad R=\frac{1}{n},
\eeq
for $A=1,2,\cdots, 5$. Here,
$G^{(n)}_A$ are $N\times N$ matrices acting on ${\cal H}_n$ and 
are given by the $n$-fold symmetric tensor products of the 
five-dimensional Euclidean gamma matrices $\Gamma_A$:
\beq
	G^{(n)}_{A}
	=\Gamma_{A}\otimes\1_4 \otimes\cdots\otimes\1_4
+\1_4 \otimes \Gamma_{A}\otimes \cdots\otimes\1_4
	+\cdots+\1_4\otimes\cdots\otimes\1_4\otimes\Gamma_{A}.
\label{def of g}
\eeq
We emphasize that though $G^{(n)}_{A}$ are represented as 
$4^n \times 4^n$ matrices, the Hilbert space is now restricted 
to ${\cal H}_n$ with dimension (\ref{dim of hn}).
As we prove in appendix
\ref{Derivation of useful relations for fuzzy S4}, 
the matrices $G^{(n)}_{A}$ satisfy the relation,
\begin{align}
\sum_A (G^{(n)}_A)^2 = n(n+4) \1_{{\cal H}_n}.
\label{GG=1}
\end{align}
The normalization factor $R$ 
is chosen so that $X_A$ gives a unit sphere in the large-$N$ limit:
\beq
\sum_A	X_A^2=\1_{{\cal H}_n} +\mathcal{O}(1/n).
\label{XX=1 for fuzzy s4}
\eeq
Below, we will see that the classical spaces defined by the 
Dirac operator and Hamiltonian indeed become the unit sphere 
in the large-$N$ limit.

\subsubsection{Classical space for fuzzy $S^4$}
\label{Classical space for fuzzy S4}
Here, we compute the classical space of fuzzy $S^4$ 
by using the Dirac operator method (See also \cite{Karczmarek:2015gda}). 
See appendix~\ref{Ham method for fuzzy s4} for the derivation using the 
Hamiltonian method.

For the configuration 
(\ref{Matrix for fuzzy S4}) of fuzzy $S^4$, the Dirac operator is given by
\begin{align}
\slashchar{D}(y) = \Gamma^A \otimes (RG^{(n)}_A -y_A).
\end{align}
This is a $4N\times 4N$ 
Hermitian matrix acting on $C^4 \otimes {\cal H}_n$.
In \cite{Karczmarek:2015gda}, 
it is shown that this Dirac operator 
has $n+2$ degenerate zero modes. In the following, we 
present these states based on a symmetry argument. 
We parametrize $y_A$ as $y_A=|y|x_A$, where 
$|y|=\sqrt{\sum_{A}y_A^2}$ and $x_A$ is the unit vector 
(\ref{vector e}) parametrized with the polar coordinates.
We consider a similarity transformation of $\slashchar{D}(y)$ 
with the unitary matrix defined in (\ref{U for fuzzy S^4}). 
Because of the relations 
(\ref{e to e0 for fuzzy s4}) and (\ref{Jan9-1}), the Dirac operator 
transforms into 
\begin{align}
U^{\dagger\otimes (n+1)}
\slashchar{D}(y)
U^{\otimes (n+1)}
 = \sum_{a=1}^4 \Gamma^a \otimes RG^{(n)}_a 
+ \Gamma^5 \otimes (RG^{(n)}_5-|y|).
\label{UDU for S4}
\end{align}
By using (\ref{gamma gamma eta eta}) and 
$\Gamma^5 | \eta_1 \rangle = | \eta_1 \rangle $, 
we can see that the $(n+1)$-fold tensor product of $|\eta_1 \rangle $ 
gives an eigenstate of (\ref{UDU for S4}) with
the eigenvalue $nR-|y|$. Thus, for $|y|=nR$, 
$U^{\otimes (n+1)}|\eta_1 \rangle^{\otimes (n+1)}$ gives a zero mode of the 
Dirac operator. Note that the vector 
$|\eta_1 \rangle^{\otimes (n+1)}$ is an element of 
${\cal H}_{n+1} \subset C^4 \otimes {\cal H}_n$.
Hence, we can consider the action of 
$SO(5)$ generators $D_{{\cal H}_{n+1}}(\Sigma_{AB})$
onto this vector. 
We notice that (\ref{UDU for S4}) commutes with the $SO(4)$ generators 
$D_{{\cal H}_{n+1}}(\Sigma_{ab})$ with $a,b =1,2,3,4$, since all the
$SO(4)$ vector indices are contracted in (\ref{UDU for S4}).
Thus, any states given by acting these generators on 
$|\eta_1 \rangle^{\otimes (n+1)}$ also give zero 
eigenstates of (\ref{UDU for S4}). In order to write down these states, 
we utilize the decomposition of the $SO(4)$ generators into the generators of $SU(2)\times SU(2)$:
\begin{align}
&J_1= -\frac{i}{2}(\Sigma_{41}+\Sigma_{23}), \;\; 
J_2= -\frac{i}{2}(\Sigma_{42}+\Sigma_{31}), \;\;
J_3= -\frac{i}{2}(\Sigma_{43}+\Sigma_{12}), \nonumber\\
&\tilde{J}_1= \frac{i}{2}(\Sigma_{41}-\Sigma_{23}), \;\; 
\tilde{J}_2= \frac{i}{2}(\Sigma_{42}-\Sigma_{31}), \;\;
\tilde{J}_3= \frac{i}{2}(\Sigma_{43}-\Sigma_{12}).
\label{su2 decomposition}
\end{align}
They satisfy 
\begin{align}
[J_i, J_j]= i \epsilon_{ijk}J_k, \;\; 
[\tilde{J}_i, \tilde{J}_j]= i \epsilon_{ijk}\tilde{J}_k, \;\;
[J_i, \tilde{J}_j]=0.
\end{align}
It is easy to see that $D_{{\cal H}_{n+1}}(\tilde{J}_i)$ are vanishing 
on $|\eta_1 \rangle^{\otimes (n+1)}$, while 
$D_{{\cal H}_{n+1}}(J_i)$ act as
\begin{align}
&D_{{\cal H}_{n+1}}(J_3)|\eta_1 \rangle^{\otimes (n+1)} = \frac{n+1}{2} 
|\eta_1 \rangle^{\otimes (n+1)}, \nonumber\\
&D_{{\cal H}_{n+1}}(J_i)^2|\eta_1 \rangle^{\otimes (n+1)} = 
\frac{(n+1)(n+3)}{4} 
|\eta_1 \rangle^{\otimes (n+1)}.
\end{align}
Hence, the state $|\eta_1 \rangle^{\otimes (n+1)}$ is the 
highest weight state under the one of the $SU(2)$ symmetries.
We use the notation $J=\frac{n+1}{2}$ for the spin of this state and 
label the highest state as 
\begin{align}
|\eta_1 \rangle^{\otimes (n+1)} = | J,J \rangle.
\end{align}
By acting $D_{{\cal H}_{n+1}}(J_-)$ on this state, 
we can obtain the other zero eigenstates of (\ref{UDU for S4}). 
By multiplying $U^{\otimes (n+1)}$ on these states, 
we finally obtain the $n+2$ degenerate zero eigenstates of 
$\slashchar{D}(y)$ as 
\begin{align}
U^{\otimes (n+1)}|J, m \rangle = 
\sqrt{\frac{(J+m)!}{(2J)!(J-m)!}}
U^{\otimes (n+1)} D_{{\cal H}_{n+1}}(J_-)^{J-m} |\eta_1 \rangle^{\otimes (n+1)},
\label{zero modes of fuzzy s4}
\end{align}
where $m= -J, -J+1, \cdots, J$. These states have the common
eigenvalue $Rn-|y|$ for the Dirac operator.

The classical space is given by the loci of zeros of
the Dirac operator as
\begin{align}
{\cal M}= \{ y \in \textbf{R}^5 | |y|=nR  \}.
\label{M for S4}
\end{align}
This is indeed $S^4$ with radius $nR=1$.
Note that this radius differs from
the naive expectation (\ref{XX=1 for fuzzy s4}) by 
$1/n$ corrections.

\subsubsection{Information metric and Berry connection for fuzzy $S^4$}
We introduce the spherical coordinate for (\ref{M for S4}) by parameterizing 
$y^A$ as 
\begin{align}
y^A = Rn x^A,
\end{align}
where $x^A$ is defined in (\ref{vector e}).
The information metric for the fuzzy $S^4$ is given by 
\beq
	\label{IM on four sphere 1}
	d^2s
	=\frac{1}{n+2}
\left(
\sum_{m=-J}^J || dU^{\otimes (n+1)} | J,m \rangle ||^2
-\sum_{m,m'=-J}^J |\langle J,m|U^{\dagger \otimes (n+1)}dU^{\otimes (n+1)}
|J,m' \rangle  |^2
\right).
\eeq
To evaluate this metric, 
let us introduce the chiral projection operators,
\beq
	P_\pm=\frac{1}{2}(\1_4\pm\Gamma_5).
\eeq
Note that the states $|J,m\rangle $ have the positive chirality,
\begin{align}
P_\pm^{\otimes (n+1)}|J,m \rangle = |J,m \rangle. 
\end{align}
We notice that the second term in (\ref{IM on four sphere 1}) 
can be written as
\begin{align}
-\frac{1}{n+2}\sum_{m,m'=-J}^J
\langle J,m| U^{\dagger \otimes (n+1)}dU^{\otimes (n+1)}
|J,m' \rangle \langle J,m' |
dU^{\dagger \otimes (n+1)} U^{\otimes (n+1)} |J,m \rangle,
\label{udu comp}
\end{align}
and $U^\dagger dU$ takes values in the $SO(5)$ Lie algebra.
Since the both sides of $U^\dagger dU$ are projected onto 
the positive chirality states in (\ref{udu comp}), 
only terms with $SO(4)$ generators survive.
Furthermore, if one decomposes $SO(4)$ to $SU(2)\times SU(2)$
as in (\ref{su2 decomposition}), 
$\tilde{J}_i$ vanish on $|J,m\rangle$. 
Thus, only the generators $J_i$ 
in $U^\dagger dU$ survive in computing (\ref{udu comp}).
For the generators $J_i$, $\sum_{m'}|J,m' \rangle \langle J,m' |$
behaves as the unit matrix. In other words, we have 
\begin{align}
\sum_{m=-J}^J|J,m \rangle \langle J,m | = P_+^{\otimes (n+1)}.
\label{completeness on s4}
\end{align}
Hence, (\ref{udu comp}) is equivalent to
\begin{align}
-\frac{1}{n+2}\sum_{m=-J}^J
\langle J,m| U^{\dagger \otimes (n+1)}dU^{\otimes (n+1)}
P_+^{\otimes (n+1)}
dU^{\dagger \otimes (n+1)} U^{\otimes (n+1)} |J,m \rangle.
\end{align}
Combining this with the first term in (\ref{IM on four sphere 1}), 
we find that the information metric is written as
\begin{align}
&ds^2 = \frac{1}{n+2}\sum_{m=-J}^J 
\langle J, m| {\cal O} |J,m \rangle, \;\;\; \;
{\cal O}= P_+dU^\dagger U P_- U^\dagger dU P_+ \otimes 
\1_4^{\otimes n}
+ \cdots, 
\label{ds2 for fuzzy s4}
\end{align}
where $\cdots$ stands for the $(n+1)$-fold symmetrization of the 
first term. 
From (\ref{Form of U}), we can explicitly compute 
$U^\dagger dU$ as
\al{
	U^{\dagger}dU
	&=\frac{1}{2}d\theta\Gamma_{45}+\frac{1}{2}d\phi\qty(\cos\theta\Gamma_{34}
	+\sin\theta\Gamma_{35}) \nonumber \\
	&\quad+\frac{1}{2}d\psi\qty{\cos\phi\Gamma_{23}
	+\sin\phi\qty(\cos\theta\Gamma_{24}+\sin\theta\Gamma_{25})} \nonumber \\
	&\quad+\frac{1}{2}d\chi\qty[\cos\psi\Gamma_{12}
	+\sin\psi\qty{\cos\phi\Gamma_{13}+\sin\phi\qty(\cos\theta\Gamma_{14}
	+\sin\theta\Gamma_{15})}].
}
This is decomposed under the chiral projection as 
\begin{align}
P_+U^\dagger dU P_+ 
= &P_+ \left\{
\frac{d\phi}{2}\cos \theta \Gamma_{34}
+\frac{d\psi}{2}(\cos\phi\Gamma_{23}+\sin\phi\cos\theta \Gamma_{24})
\right. \nonumber\\
& \left. +\frac{d\chi}{2}(\cos\psi\Gamma_{12}+\sin\psi\cos\phi \Gamma_{13}+
\sin\psi\sin\phi\cos\theta\Gamma_{14})
\right\} P_+, \nonumber\\
P_-U^\dagger dU P_+ 
= &P_- \left\{
\frac{d\theta}{2}\Gamma_{45}
+\frac{d\phi}{2}\sin\theta\Gamma_{35}
+\frac{d\psi}{2}\sin\phi\sin\theta\Gamma_{25}
+\frac{d\chi}{2} \sin\psi\sin\phi\sin\theta \Gamma_{15}
\right\} P_+.
\label{decomposing UdU}
\end{align}
By substituting this into (\ref{ds2 for fuzzy s4}), finally 
we find that 
\beq
ds^2
=\frac{n+1}{4}
\qty(d\theta^2+\sin^2\theta\,d\phi^2+\sin^2\theta\sin^2\phi\,d\psi^2
+\sin^2\theta\sin^2\phi\sin^2\psi\,d\chi^2).
\eeq
This is the standard round metric for $S^4$.

Next, we calculate the Berry connection for fuzzy $S^4$, which 
is defined by 
\begin{align}
A_{mm'}= -i \langle J,m | U^{\dagger \otimes (n+1)}
dU^{\otimes(n+1)}|J,m' \rangle.
\end{align}
By using (\ref{su2 decomposition}) and 
(\ref{decomposing UdU}), we find that the Berry connection is 
given by
\begin{align}
A_{mm'}=&(\cos\phi d\psi-\sin\psi\sin\phi\cos\theta d\chi)
D_{{\cal H}_{n+1}}(J_1)_{mm'} \nonumber\\
&-(\sin\phi\cos\theta d\psi + \sin\psi \cos\phi d\chi)
D_{{\cal H}_{n+1}}(J_2)_{mm'} \nonumber\\
&-(\cos\theta d\phi - \cos\psi d\chi)
D_{{\cal H}_{n+1}}(J_3)_{mm'}.
\label{berry for s4}
\end{align}
Let us also calculate the field strength.
Introducing the
matrix notation, $A:= \sum_{i=1}^3A_{i}D_{{\cal H}_{n+1}}(J_i)$,
the field strength is given by 
$F= \sum_{i=1}^3F_{i}D_{{\cal H}_{n+1}}(J_i)$, where
\bea
	F^a &=d A^a - \frac{1}{2} \epsilon^{abc} A^b\wedge A^c ,
\ena
Straightforward calculation gives
\begin{align}
&F^1 = 
-\sin\phi\sin^2\theta d\phi \wedge d\psi
+\sin\psi\sin\phi\sin\theta d\theta \wedge d\chi,
\nonumber\\
&F^2 =
\sin\phi\sin\theta d\theta \wedge d\psi
+\sin\psi\sin\phi\sin^2\theta d\phi \wedge d\chi,
\nonumber\\
&F^3 =
\sin\theta d\theta\wedge d\phi
-\sin\psi\sin^2\phi\sin^2\theta d\psi\wedge d\chi.
\end{align}
We can also show that this configuration is self-dual.
Taking a square of the field strength, we obtain
\bea
	{F}^1 \wedge {F}^1+ {F}^2 \wedge {F}^2+ {F}^3 \wedge {F}^3
	=6 \sin^3\theta\sin^2\phi\sin\psi (d\chi \wedge d\psi \wedge d\phi\wedge d \theta ).
\ena
The right-hand side is just a volume form on $S^4$.
This configuration is known as the $SU(2)$ Yang monopole on $S^4$.
The instanton number (The second Chern class) 
is given by the matrix size:
\begin{align}
\frac{1}{8\pi^2} \int {\rm Tr}_{{\cal H}_{n+1}}(F\wedge F) =N. 
\label{2nd Chern class}
\end{align}

\section{Berezin-Toeplitz quantization}
\label{Berezin-Toeplitz quantization}
In this section, we show that the matrix configurations of fuzzy 
$S^2$ and fuzzy $S^4$ can be regarded as the images of the 
Berezin-Toeplitz quantization map.

\subsection{Review of Berezin-Toeplitz quantization}
We first give a brief review of the Berezin-Toeplitz quantization map on 
spin-C manifold. 
We consider a Euclidean compact spin-C manifold ${\cal M}$ with a 
Riemannian metric $g$ and a spinor bundle on ${\cal M}$.
We assume that the gauge group is $U(k)$ and 
the spinors shall belong to 
the representation ${\cal R}$ of the gauge group. 
We define the Dirac operator as usual as
\begin{align}
{\cal D}\!\!\!\!/ = \Gamma^A e_A^\mu 
(\partial_{\mu}+\frac{1}{4}\omega_{\mu BC}\Gamma^{BC}-iA_\mu ),
\label{contiuum dirac op}
\end{align}
where $A$, $e$ and $\omega$ are the gauge connection, 
vielbein and spin connection, respectively.
By using the invariant measure defined from the metric $g$, 
we can define the inner product of sections. We denote this inner 
product as $(\psi, \psi')$. 

Because of the index theorem, the kernel of the Dirac operator
(\ref{contiuum dirac op}) forms a finite dimensional vector space.
The dimension of this vector space is related to the Chern numbers
of $A$ as well as the representation ${\cal R}$ of spinors. 
We denote this dimension by $N$. 
Let $\{\psi_i | i=1,2,\cdots,N\}$ be an orthonormal basis of 
${\rm Ker}{\cal D}\!\!\!\!/$ satisfying $(\psi_i, \psi_j)=\delta_{ij}$. 
Multiplying a function $f\in C^\infty(\cal M) $ on 
$\psi_i$ gives another spinor, which in general does not 
belong to ${\rm Ker}{\cal D}\!\!\!\!/ \;$ and can be expanded in terms of 
the eigen functions of ${\cal D}\!\!\!\!/ \;$ as
\begin{align}
f \psi_i = \sum_{j=1}^N \hat{f}_{ij} \psi_j  + \cdots.
\end{align}
$\hat{f}_{ij}$ are constants (coefficients of $\psi_j$ in this expansion), and 
$\cdots$ represents the part which takes values in the orthogonal 
complement of ${\rm Ker}{\cal D}\!\!\!\!/$. 
The coefficients $\hat{f}_{ij}$ can be extracted as 
\begin{align}
\hat{f}_{ij} = (\psi_j, f\psi_i).
\label{def of Toeplitz map}
\end{align}
Since $\{ \hat{f}_{ij} \}$ is just a constant $N\times N$ matrix, 
this construction can be seen as a mapping from a function $f$ to 
an $N\times N$ matrix. 
This is the Berezin-Toeplitz quantization map.
The matrix $\hat{f}$ is called the Toeplitz operator of $f$.

\subsection{Berezin-Toeplitz quantization for fuzzy $S^2$}
Here, we will show that the matrix configuration (\ref{X for fuzzy s2}) 
is equal to the Toeplitz operator 
of the standard embedding function $S^2 \rightarrow \textbf{R}^3$.

We first show that the zero eigenstate $|\psi_J \rangle $ 
in (\ref{eigenstates of fuzzy S2}) also gives
a zero eigenstate of the continuum 
Dirac operator (\ref{contiuum dirac op}). 
In order to fix the basis of the 2-component spinors, 
we make a local Lorentz transformation and consider 
\begin{align}
|\psi_J \rangle \! \rangle = U_2^{\dagger}\otimes 1_N |\psi_J \rangle
= 
\left(
\begin{array}{c}
U_N |J,J \rangle \\
0 \\
\end{array}
\right).
\label{local lorentz on s2}
\end{align}
Note that $|\psi_J \rangle \! \rangle$ contains only 
the positive chirality component $|1/2,+ \rangle$, and this is 
written as the upper component in the last expression in 
(\ref{local lorentz on s2}).
By using the vielbein and spin connection in
appendix~\ref{Spin connections on s2 and s4}, 
we can write the covariant derivatives 
$\nabla_a= e_a^\mu (\partial_\mu +\frac{1}{4}\omega_{\mu bc}\sigma^{bc})$ 
explicitly as
\begin{align}
&\nabla_+ = \frac{1+|z|^2}{r}
\left(
\partial_z + \frac{\bar{z}}{2(1+|z|^2)} \sigma^3
\right), \nonumber\\
&\nabla_- = \frac{1+|z|^2}{r}
\left(
\partial_{\bar{z}} - \frac{z}{2(1+|z|^2)} \sigma^3
\right),
\end{align}
where $r$ is the radius of the sphere.
The actions of these operators on $|\psi_J \rangle \! \rangle$
follow from (\ref{dpsi}) as
\begin{align}
&\nabla_+ | \psi_J \rangle \! \rangle =
\frac{(J+1/2)\bar{z}}{r} | \psi_J \rangle \! \rangle 
+\frac{\sqrt{2J}}{r} 
\left(
\begin{array}{c}
U_N |J,J-1 \rangle \\
0 \\
\end{array}
\right), \nonumber\\
&\nabla_- | \psi_J \rangle \! \rangle =
-\frac{(J+1/2)z}{r} | \psi_J \rangle \! \rangle.
\label{nabla psi}
\end{align}
Note that the first terms in these expressions are just the Berry 
connections multiplied by the inverses of vielbein, $e_a^\mu A_\mu$. 
From (\ref{nabla psi}), we find that $| \psi_J \rangle \! \rangle$ satisfies
\begin{align}
{\cal D}\!\!\!\!/ \; | \psi_J \rangle \! \rangle =
\sigma^a(\nabla_a -ie_a^\mu A_\mu ) | \psi_J \rangle \! \rangle =0.
\end{align}
Thus, $| \psi_J \rangle \! \rangle$ is a zero eigenstate
of ${\cal D}\!\!\!\!/ \;$. 
There are $N$ independent components in $| \psi_J \rangle \! \rangle$. 
Introducing the basis $|i\rangle$$(i=1,2,\cdots,N)$ of the 
$N$-dimensional vector space, we thus find $N$ 
zero modes of ${\cal D}\!\!\!\!/ \;$:
\begin{align}
\psi_i = 
\left(
\begin{array}{c}
\langle i | U_N |J,J \rangle \\
0 \\
\end{array}
\right), \;\;\;  {\cal D}\!\!\!\!/ \; \psi_i =0 .
\end{align}

By using the information metric $g$, we can define the 
standard inner product for spinors. 
In the following calculations, we use the formulas, 
\begin{align}
\int d\Omega_2 \frac{|z|^{2B}}{(1+|z|^2)^A}
=4 \pi \frac{(A-B)! B!}{(A+1)!}
\end{align}
and
\begin{align}
U_N|J,J \rangle = 
\frac{1}{(1+|z|^2)^J}\sum_{r=-J}^J z^{J-r}
\left(
\begin{array}{c}
2J \\
J+r \\
\end{array}
\right)^{1/2} | J, r \rangle,
\end{align}
where $d\Omega_2=\sin\theta d\theta \wedge d \phi = 
\frac{2i dz\wedge d\bar{z}}{(1+|z|^2)^2} $ is the 
volume form of the unit sphere satisfying 
$\int d\Omega_2 =4\pi$.
From these formulas, we can 
easily show that\footnote{
The equation (\ref{unity}) can also be obtained easily from the symmetry 
argument: The integration over $S^2$ only produces 
rotationally invariant tensors on $S^2$. From the structure of indices, 
the integration of $U_N |J,J \rangle \langle J,J | U_N^{\dagger}$
turns out to be proportional to the identity matrix. The proportionality 
constant is fixed by taking the trace.
}
\begin{align}
\frac{N}{4\pi}\int d\Omega_2 U_N |J,J \rangle \langle J,J | U_N^{\dagger}
=\1_N.
\label{unity}
\end{align}
This implies that $\psi_i$ are orthonormal under the 
inner product given by the information metric.
From the index theorem, it also follows that 
the dimension of ${\rm Ker}{\cal D}\!\!\!\!/ \; $ is 
equal to $N$. Thus, $\psi_i$ form an orthonormal basis of 
${\rm Ker}{\cal D}\!\!\!\!/ \; $.

The Toeplitz operator for a function $f\in C^{\infty} (S^2)$
is given by 
\begin{align}
\hat{f}_{ij}= 
(\psi_j, f\psi_i)= 
\frac{N}{4\pi }
\int d\Omega_2
\langle i | U_N |J,J \rangle f \langle J,J | U_N^{\dagger} | j \rangle.
\label{fij for s2}
\end{align}
The formula (\ref{unity})
also implies that the image of the unit constant function 
on $S^2$ is given by the identity matrix. 
Similarly, we can compute the images of the embedding functions
$x^i $ $(i=1,2,3)$ defined in  (\ref{unit vector on s2}).
We find that they are mapped to
\begin{align}
\hat{x}^i = \frac{1}{J+1}L^i,
\end{align} 
for $i=1,2,3$. This is just the matrix configuration 
of fuzzy $S^2$ up to the overall constant.  
Thus, the matrix configuration of the fuzzy $S^2$ can be regarded
as the Toeplitz operator of the embedding function 
$S^2 \rightarrow \textbf{R}^3$.

The Toeplitz quantization map also induces mappings for derivatives 
and integrals on $S^2$. For example, the mapping 
rule for  integrals on $S^2$ can be obtained by taking 
the trace of (\ref{fij for s2}):
\begin{align}
{\rm Tr} \hat{f} = \frac{N}{4\pi} \int d\Omega_2 f.
\end{align}
Thus, integrals are mapped to traces. Similarly, we can derive 
the mapping rule for the Laplace operator on $S^2$ as 
\begin{align}
(\hat{\Delta f})_{ij} = -\frac{1}{r^2}[L_k,[L_k,\hat{f}]]_{ij}.
\label{mapping delta s2}
\end{align}
See appendix \ref{Laplacians on S^2} for derivation.

\subsection{Berezin-Toeplitz quantization for fuzzy $S^4$}

Next, we show that the matrix configuration (\ref{Matrix for fuzzy S4}) 
is equal to the Toeplitz operator 
of the standard embedding function $S^4 \rightarrow \textbf{R}^5$.
We also obtain the mapping rule for the Laplacian on $S^4$.

We perform a local Lorentz transformation of the zero eigenstates
(\ref{zero modes of fuzzy s4}) and consider
\begin{align}
|\psi_{Jm}\rangle \! \rangle = \1_4 \otimes U^{\otimes n}|J, m \rangle.
\end{align}
Recall that $|J, m \rangle$ is an element of 
${\cal H}_{n+1} \subset C^4 \times {\cal H}_n$ and 
can be written as a sum of tensor products of elements in 
$C^4$ and ${\cal H}_n$. This decomposition is given by
\begin{align}
|J, m \rangle = \sum_{s=-\frac{1}{2}}^{\frac{1}{2}} 
\sum_{\gamma=-J+\frac{1}{2}}^{J-\frac{1}{2}}
C^{Jm}_{\frac{1}{2}s J-\frac{1}{2} \gamma} 
| 1/2, s  \rangle \otimes
|J-1/2, \gamma \rangle,
\end{align}
where $C^{c \gamma}_{a\alpha b\beta}$ is the Clebsh-Gordan coefficient of 
$SU(2)$. In terms of this expression, 
$|\psi_{Jm}\rangle \! \rangle$ can also be written as
\begin{align}
|\psi_{Jm}\rangle \! \rangle 
= \sum_{s=-\frac{1}{2}}^{\frac{1}{2}} 
\sum_{\gamma=-J+\frac{1}{2}}^{J-\frac{1}{2}}
C^{Jm}_{\frac{1}{2}s J-\frac{1}{2} \gamma} 
| 1/2, s  \rangle \otimes
( U^{\otimes n}|J-1/2, \gamma \rangle ). 
\label{psi another expression}
\end{align}
Note that this vector has the positive chirality with respect to 
$\Gamma_5$: 
\begin{align}
\Gamma_5 \otimes \1_4^{\otimes n} |\psi_{Jm}\rangle \! \rangle =
|\psi_{Jm}\rangle \! \rangle.
\end{align}
Below, we will show that 
$|\psi_{Jm}\rangle \! \rangle$ is a zero eigenvector of
the differential Dirac operator (\ref{contiuum dirac op}).
Here, the gauge field is given by the Berry connection 
(\ref{berry for s4}) and the representation of the 
gauge group is the spin $J=\frac{n+1}{2}$ representation of $SU(2)$. 
The vielbein and the spin connection are given in 
appendix~\ref{Spin connections on s2 and s4}.

Let us first consider the action of the covariant derivative 
$\nabla_a= e_a^\mu (\partial_\mu +\frac{1}{4}\omega_{\mu bc}\Gamma^{bc})$
without the gauge connection. 
By comparing (\ref{decomposing UdU}) and (\ref{omega for s4}), 
we find that the spin connection has the following relation:
\begin{align}
\frac{1}{4}\sum_{a,b=1}^4\omega_{ab}\Gamma^{ab} P_+ = P_+U^\dagger dU P_+.
\end{align}
By using this relation, we obtain
\begin{align}
\Gamma^a \nabla_a  |\psi_{Jm}\rangle \! \rangle
= (\Gamma^a P_+ \otimes U^{\otimes n})
(U^\dagger \partial_a U \otimes \1_4^{\otimes n} + \cdots)
|J,m \rangle,
\label{tochyu}
\end{align}
where $\partial_a = e_a^\mu \partial_\mu$ and 
$\cdots $ stands for the symmetrization of the first term.
In the symmetrization of $U^\dagger \partial_a U$, we insert 
$\1_4=P_++P_-$ in front of each $U^\dagger \partial_a U$.
Then, the terms containing $P_+$ in these insertions can be calculated as
\begin{align}
&(\Gamma^a P_+ \otimes (UP_+)^{\otimes n} )
(U^\dagger \partial_a U \otimes \1_4^{\otimes n} + \cdots)
|J,m \rangle  
\nonumber\\
=& \sum_{m'=-J}^J (\Gamma^a \otimes U^{\otimes n} )
|J,m' \rangle \langle J,m' |
 (U^\dagger \partial_a U \otimes \1_4^{\otimes n} + \cdots)
|J,m \rangle
\nonumber\\
=& \sum_{m'=-J}^J (\Gamma^a \otimes \1_4^{\otimes n} ) 
|\psi_{Jm'}\rangle \! \rangle  (iA_a)_{m'm},
\end{align}
where the second line follows from (\ref{completeness on s4}).
Thus, this contribution gives the Berry connection. 
On the other hand, the terms containing $P_-$ can be calculated as
\begin{align}
&(\Gamma^a P_+ \otimes U^{\otimes n})
\times \1_4 \otimes (P_- U^\dagger \partial_a U P_+ \otimes 
\1_4^{\otimes (n-1)}  + \cdots)
|J,m \rangle
\nonumber\\
=& 
\frac{1}{2r} (\1_4 \otimes U^{\otimes n})
\times \Gamma^a \otimes (\Gamma_a \otimes 
\1_4^{\otimes (n-1)}  + \cdots)
|J,m \rangle 
\nonumber\\
=& 
\frac{1}{2r} (\1_4 \otimes U^{\otimes n})
(\Gamma^a \otimes G_a^{(n)}) 
|J,m \rangle,
\end{align}
where we used (\ref{decomposing UdU})
and (\ref{vielbein on s4}) to obtain the second line.
Note that the last expression is vanishing as we saw in 
section~\ref{Classical space for fuzzy S4}. 
Thus, combining these calculations, 
we find that 
$|\psi_{Jm}\rangle \! \rangle $ gives a zero eigenvector of 
the gauge covariant Dirac operator (\ref{contiuum dirac op}):
\begin{align}
{\cal D}\!\!\!\!/ \; 
|\psi_{Jm}\rangle \! \rangle =0,
\label{Dpsi=0 for s4}
\end{align}
where the gauge field acts as 
$A |\psi_{Jm}\rangle \! \rangle =
 \sum_{m'}|\psi_{Jm'}\rangle \! \rangle (A)_{m'm}$.

Let $\{ |i \rangle  | i=1,2,\cdots, N \}$ be any orthonormal 
basis of ${\cal H}_n$.
By multiplying $\1_4 \otimes \langle i | $ 
to the state (\ref{psi another expression}), we obtain 
\begin{align}
\psi_i^{Jm} := \frac{\sqrt{N}}{n+1}
\sum_{s=-\frac{1}{2}}^{\frac{1}{2}} 
\sum_{\gamma=-J+\frac{1}{2}}^{J-\frac{1}{2}}
C^{Jm}_{\frac{1}{2}s J-\frac{1}{2} \gamma} 
| 1/2, s  \rangle 
\langle i | U^{\otimes n} |J-1/2, \gamma \rangle . 
\label{zero eigen spinors on s4}
\end{align}
$\psi_i^{Jm} $ are $N$ spinors on $S^4$, which are 
also elements of ${\rm Ker}{\cal D}\!\!\!\!/ \; $ because of 
(\ref{Dpsi=0 for s4}).
We introduce a gauge invariant inner product between these spinors by 
\begin{align}
(\psi_i, \psi_j)
= \frac{1}{n+2}\sum_{m=-J}^J \frac{3 (n+1)^2}{8\pi^2}
\int d\Omega_4 (\psi_i^{Jm})^{\dagger} \cdot \psi_j^{Jm},
\label{def of inner product on s4}
\end{align}
where the dot $\cdot$ between $\psi$'s stands for the contraction 
of the spinor indices, and $d\Omega_4$ is the volume 
form of the unit $S^4$ normalized as $\int d\Omega_4 = \frac{8\pi^2}{3}$. 
We multiplied the factor $(n+1)^2$ so that the integration 
measure becomes proportional to the invariant measure made 
of the information metric.

Let us calculate $(\psi_i, \psi_j)$.
By using  (\ref{zero eigen spinors on s4}), we obtain
\begin{align}
\sum_m (\psi_i^{Jm})^{\dagger} \cdot \psi_j^{Jm}
= \frac{N}{(n+1)^2} \sum_{m,x,\gamma,\gamma'}
C^{Jm}_{\frac{1}{2}sJ-\frac{1}{2}\gamma}
C^{Jm}_{\frac{1}{2}sJ-\frac{1}{2}\gamma' }
\langle j |U^{\otimes n} | J-1/2, \gamma' \rangle
\langle J-1/2, \gamma | U^{\dagger \otimes n} | i \rangle.
\end{align}
By using the summation formula of the Clebsh-Gordan coefficients, 
\begin{align}
\sum_{\alpha, \gamma}
C^{c \gamma}_{a \alpha b \beta}
C^{c \gamma}_{a \alpha b' \beta'}
= \frac{2c+1}{2b+1} \delta_{bb'}\delta_{\gamma \gamma '},
\end{align}
we obtain
\begin{align}
\sum_m \int d\Omega_4 (\psi_i^{Jm})^{\dagger} \cdot \psi_j^{Jm}
=\frac{N}{(n+1)^2}\frac{2J+1}{2J}
\int d\Omega_4  
\langle j | (U P_+ U^{\dagger})^{\otimes n} | i \rangle.
\label{int psi psi}
\end{align}
To obtain the the last expression, we also used the fact 
that $| J-1/2, \gamma \rangle$ forms a complete 
basis of ${\cal H}^+_n$ and satisfies 
\begin{align}
\sum_{\gamma} | J-1/2, \gamma \rangle
\langle J-1/2, \gamma | = P_+^{\otimes n}.
\end{align}
Finally, by using (\ref{1st eq}), we find that 
(\ref{int psi psi}) is given by $\delta_{ij}$
multiplied by a constant factor.
Substituting this result into 
(\ref{def of inner product on s4}), we find that 
\begin{align}
(\psi_i, \psi_j) = \delta_{ij}.
\label{orthonormal}
\end{align}
Namely, $\psi_i^{Jm}$ are orthonormal under this inner product.
Note that, from the index theorem with the second Chern class 
(\ref{2nd Chern class}), the dimension of 
${\rm Ker}{\cal D}\!\!\!\!/ \; $ is equal to $N$. 
Thus, $\{ \psi_i^{Jm} | i=1,2, \cdots, N \}$ gives
a complete basis of  ${\rm Ker}{\cal D}\!\!\!\!/ \; $.

We then consider the Toeplitz quantization map 
(\ref{def of Toeplitz map}) for fuzzy $S^4$. 
Note that the orthonormal relation (\ref{orthonormal}) 
implies that the image of the unit constant function on $S^4$ is 
equal to the identity matrix $\1_{{\cal H}_n}$. 
In this paper, we assume for simplicity that the function $f$
is gauge singlet, namely, it is proportional to
$\delta_{mm'}$. 
In this case, (\ref{def of Toeplitz map}) can be 
written more explicitly as
\begin{align}
f_{ij}= \frac{3}{8\pi^2}\frac{N}{n+1}
\int d\Omega_4 \; f \; \langle i | 
(UP_+U^\dagger )^{\otimes n} |j \rangle.
\label{fij for fuzzy s4}
\end{align}
Let us consider the case in which 
$f$ is the embedding function 
$x^A$ defined in (\ref{vector e}).
By using the formula (\ref{2nd eq}), we find that 
the image of this embedding function is given as
\begin{align}
\hat{x}^A_{ij} = \frac{1}{n+4} \langle i |G_A^{(n)} | j\rangle.
\end{align}
The right-hand side is just 
the matrix configuration of fuzzy $S^4$.
Thus, we find that the configuration of fuzzy $S^4$ can 
be obtained as the Toeplitz operator of 
the embedding function $S^4 \rightarrow \textbf{R}^5$.

As for the case of $S^2$, we can obtain the mapping rules 
for integrals and the Laplace operator on $S^4$.
By taking the trace of (\ref{fij for fuzzy s4}), 
we obtain
\begin{align}
{\rm Tr}\hat{f} = \frac{3N}{8\pi^2} \int d\Omega_4 f.
\end{align}
Thus, integrals are mapped to traces of matrices.
Similarly, the image of the Laplace operator on $S^4$ is 
given by 
\begin{align}
(\hat{\Delta f})_{ij} = -\frac{1}{4 r^2}[G_A^{(n)},[G_A^{(n)},\hat{f}]]_{ij}.
\label{mapping delta s4}
\end{align}
See appendix \ref{Laplacians on S^4} for derivation.

\section{Summary and discussion}
\label{Summary and discussion}
In this paper, we developed the notion of the information metric 
and Berry connection in the context of the matrix geometry. 
These geometric objects can be defined purely 
from given matrix configurations and are very useful in 
characterizing the geometry of matrices.
We utilized these objects to see that the well-known 
matrix configurations of fuzzy $S^2$ and fuzzy $S^4$ 
can be viewed in a unified manner as the Toeplitz operators of 
the embedding functions 
$S^n \rightarrow \textbf{R}^{n+1}$ $(n=2,4)$.
Based on this result, we also obtained mapping rules 
for the Laplacian on these 
spaces and found that in both cases, the Laplacian is 
realized as the matrix Laplacian, $[X^\mu, [X^\mu, \;\; ]]$.

The fuzzy $S^2$ is the Toeplitz quantization such that 
the gauge group is $U(1)$ and the monopole charge of 
the connection 1-form is related to the matrix size $N$. 
The large-$N$ limit corresponds to the limit of large
monopole charge. 
On the other hand, we found that the Toeplitz quantization 
map for fuzzy $S^4$ has a very different structure. 
The gauge group is non-Abelian and only an 
$SU(2)$ subgroup has nontrivial gauge connection, which 
takes the form of the Yang-monopole on $S^4$.
The Yang-monopole configuration has a fixed instanton number,
which is equal to $1$. Thus, the topological charge 
does not correspond to the matrix size unlike the case 
of fuzzy $S^2$. 
Instead, the spinors in the quantization map belong  
to the spin-$J$ representation of the $SU(2)$ subgroup 
and this spin $J$ is ultimately related to the
matrix size of fuzzy $S^4$. 
Thus, the large $N$ limit is not the limit of large 
instanton number but the limit of the large 
representation space of spinors.

It would be an interesting problem to construct a 
different Toeplitz quantization on $S^4$ such that 
the representation of spinors are fixed but the instanton numbers
are given as an increasing sequence. 
Such map would give a new description of fuzzy $S^4$.

What we argued in this paper can be understood as an  
inverse problem of 
constructing the Berezin-Toeplitz quantization. 
In the Berezin-Toeplitz quantization,
the matrices (Toeplitz operators) are 
constructed from the geometric structures such as the metric and 
gauge field, while we 
constructed the information metric and 
Berry connection from the given matrices. 
For the case of fuzzy $S^2$ and fuzzy $S^4$, 
we showed that the $N$-dimensional vector spaces on 
which the matrices $X^\mu$ are acting are indeed identified with the 
kernel of (differential) Dirac operators, and the associated Berezin-Toeplitz 
quantization produce $X^\mu$ as the Toeplitz operators. 
This means that our construction indeed
gives a solution of the inverse problem. 
Though we have checked this statement only for $S^2$ and $S^4$ in this 
paper, extending this study to more general cases should be important in 
understanding the geometry of matrices.

The use of the information metric and Berry connection
will not be limited only to the same kind of problems of 
the Berezin-Toeplitz quantization that we considered in this paper.
For example, by embedding our setup into systems with D-branes as 
considered in \cite{TachyonA, TachyonB}, 
the Berry connection will be identified with 
the gauge field on D-branes. 
Through the dualities considered in \cite{Seiberg:1999vs}, 
it will be possible to 
understand the Seiberg-Witten map for the Berry connection for 
generic configurations of D-branes. 
It will also be interesting to see the relation between 
our findings in this paper and 
some recent attempts to construct gravitational theories 
from matrix models \cite{Kaneko:2017zeo,Steinacker:2016vgf,Hanada:2005vr}.

\section*{Acknowledgments}
We thank S. Terashima for valuable discussions on the Toeplitz quantization.
The work of G. I. was supported, in part, 
by Program to Disseminate Tenure Tracking System, 
MEXT, Japan and by KAKENHI (16K17679).

\appendix
\section{Injectivity of $\rho$ and $d\rho$}
\label{Injectivity of rho and drho}
In this appendix, we prove the injectivity of $\rho$ and $d\rho$.

We first prove the injectivity of $\rho$ by contradiction. 
For $y\neq y'$ $(y, y' \in {\cal M})$, suppose that $\rho(y)=\rho(y')$.
Since the density matrix is made of zero modes of the Dirac operator, 
we have
\bea
	\slashchar{D} (y)\rho(y)=0.	\label{Rhoinj1}
\ena
From the assumption, we also have
\bea
	\slashchar{D} (y')\rho(y')=\slashchar{D} (y')\rho(y)=0. \label{Rhoinj2}
\ena
Subtracting \eqref{Rhoinj1} from \eqref{Rhoinj2}, we have
\bea
	\left(\Gamma^\mu\otimes (y_\mu-{y'}_\mu)\right)\rho(y)=0.
\ena
Similarly, from the right action of the Dirac operator, we also obtain
\bea
	\rho(y)\left(\Gamma^\mu\otimes (y_\mu-{y'}_\mu)\right)=0.
\ena
Then, we find that
\bea
	0
	&=\rho(y)\left(\Gamma^\mu\otimes (y_\mu-{y'}_\mu) \right)\left(\Gamma^\nu\otimes (y_\nu-{y'}_\nu)\right)\rho(y)\nonumber \\
	&=\frac{1}{2}\rho(y)\left(\{\Gamma^\mu,\Gamma^\nu\}\otimes (y_\mu-{y'}_\mu) (y_\nu-{y'}_\nu)\right)\rho(y)\nonumber \\
	&=(y-{y'})^2 \rho^2(y).
\ena
As we assumed $y\neq y'$, it follows that $\rho(y)=0$.
This contradicts with ${\rm Tr}\rho(y)=1$.
Hence we conclude that $\rho(y)\neq\rho(y')$ for $y\neq y'$,
which means that the map $\rho$ is injective. 

Next, we show the injectivity of the differential $d\rho$. 
Let $c^\mu(y) \partial_\mu$ be a tangent vector field  on ${\cal M}$ 
(i.e. $c^\mu$ has only tangential components along ${\cal M}$). 
We will show below that if $c^\mu \p_\mu \rho =0$, $c^\mu$ is vanishing.
This is nothing but the injectivity of $d\rho$.
Assuming $c^\mu \p_\mu \rho=0$ on ${\cal M}$, we have
\bea
	0
	&=c^\mu ( \p_\mu \rho (y))\ket[0,a,y] \nonumber\\
	&= \frac{c^\mu}{k} \left(\p_\mu \sum_{b=1}^k
\ket[0,b,y]\bra[0,b,y]\right)\ket[0,a,y] \nonumber\\
	&=\frac{c^\mu}{k}\left(1-\sum_{b=1}^k \ket[0,b,y]\bra[0,b,y]\right) 
\p_\mu\ket[0,a,y] \nonumber\\
	&=\frac{c^\mu}{k}\sum_{n\neq0}\ket[n,y]\bra[n,y]\p_\mu\ket[0,a,y].
\ena
As $\{\ket[n,y]\}$ is linearly independent, we find that
\bea
	c^\mu\bra[n,y]\p_\mu\ket[0,a,y]=0	\text{ \  for \ } n\neq 0.	\label{3.19}
\ena
From the relation, 
\begin{align}
0=c^\mu \p_\mu\left(\slashchar{D}(y)\ket[0,a,y]\right)
=-c^\mu \Gamma_\mu \ket[0,a,y]+\slashchar{D}(y)c^\mu 
\p_\mu\ket[0,a,y],
\label{relation 1}
\end{align}
it follows that
$c^\mu \bra[n,y]\p_\mu\ket[0,a,y]=c^\mu \bra[n,y]\Gamma_\mu \ket[0,a,y]/E_n$
for $n\neq 0$.
Thus, \eqref{3.19} is equivalent to
\bea
	c^\mu\bra[n,y]\Gamma_\mu\ket[0,a,y]=0	\text{ \  for \ } n\neq 0.
\label{ngamma0}
\ena
By acting $\langle 0, b, y |$ 
on the equation (\ref{relation 1}), we also obtain
\begin{align}
c^\mu \bra[0,b,y] \Gamma_\mu\ket[0,a,y]=0.
\label{0gamma0}
\end{align}
The relations (\ref{ngamma0}) and (\ref{0gamma0}) lead to
\bea
	c^\mu \Gamma_\mu\ket[0,a,y]=0.
\ena
By using this equation, we can calculate as
\bea
0	&=   \bar c^\mu c^\nu \bra[0,a,y]\Gamma_\mu\Gamma_\nu\ket[0,b,y] \nonumber \\
	&=   |c|^2 \langle 0,a,y\ket[0,b,y] \nonumber \\
	&=   |c|^2 \delta_{ab}.
\ena
This clearly shows that $c^\mu =0$. Thus, we have shown that 
$d\rho$ is an injective map.

\section{Spectrum of Dirac operator for fuzzy $S^2$ \label{ApE2}}
\label{Spectrum of Dirac Operator for Fuzzy S2}

In this appendix, we analyze the spectrum of the Dirac operator 
for the fuzzy $S^2$.
We first notice that the Dirac operator 
(\ref{Dirac operator for fuzzy S2}) satisfies
\bea
	\slashchar{D}^2(y) +R \slashchar{D}(y)
	&=(R^2J(J+1)+|y|^2) \1_2\otimes \1_N  - 2R 
\left( 
y^i \frac{\sigma_i}{2} \otimes \1_N +
\1_2\otimes y^iL_i 
\right).
\ena
Consider the operators
\begin{align}
{\cal O}_1(y) = 
y^i \frac{\sigma_i}{2} \otimes \1_N,  \;\;\;
{\cal O}_2(y) = \1_2\otimes y^iL_i.
\end{align}
Since the operators ${\cal O}_1(y)$, ${\cal O}_2(y)$ and 
$\slashchar{D}^2(y) +R \slashchar{D}(y)$ mutually commute,
they can be simultaneously diagonalized.
Thus, the eigenvalue problem of $\slashchar{D}^2(y) +R \slashchar{D}(y)$
is reduced to finding the eigenstates of 
${\cal O}_1(y)$ and ${\cal O}_2(y)$. 

The eigenstates of $y^iL_i$ can be found as follows. 
Consider the unitary similarity transformation (\ref{U undo})
which produces the rotation of the vector index. 
We consider the rotation matrix (\ref{explicit form of Lambda}) with 
$\alpha = \phi$. In this case, $U$ is explicitly given by
(\ref{Form of U}) or equivalently by (\ref{form of U 2}).
Under this similarity transformation, $y^iL_i$ transforms as
\begin{align}
y^i U^\dagger L_i U = y^i (\Lambda^{-1})_i{}^j L_j 
=|y| L_3.
\end{align}
This implies that the eigenstates of $y^iL_i$ are give by 
$U | J, m \rangle$, where $| J, m \rangle$ is the 
standard basis defined in (\ref{def of states}).
Note that $U$ depends only on the angular variables for $y$.

Diagonalizing  $y^i \frac{\sigma_i}{2}$, which appears in ${\cal O}_1(y)$,  
is just the spacial case 
of the above argument such that the dimension of the representation 
is equal to 2. Thus, its eigenstates are given by 
$U | 1/2, \pm \rangle$. 
Thus, the simultaneous eigenstates of ${\cal O}_1(y)$ and 
${\cal O}_2(y)$ are given by 
\begin{align}
U_2(y) | 1/2, \pm \rangle \otimes U_N(y)|J,m \rangle,
\label{eigenstates of o1 and o2}
\end{align}
where the subscripts of $U_2$ and $U_N$ just stand for the dimensions
of the representation spaces on which they are acting.

(\ref{eigenstates of o1 and o2}) gives the eigenstates of 
$\slashchar{D}^2(y) +R \slashchar{D}(y)$. For each eigenstate, 
the eigenvalue of $\slashchar{D}^2(y) +R \slashchar{D}(y)$ is
given by 
\begin{align}
R^2 J(J+1) +|y|^2 -2R|y|(m\pm 1/2 ).
\end{align}
Note that the states 
$U_2(y) | 1/2, + \rangle \otimes U_N(y)|J,J \rangle$
and 
$U_2(y) | 1/2, - \rangle \otimes U_N(y)|J,-J \rangle$
are not degenerate but the other states are doubly degenerate.

The non-degenerate eigenstates of 
$\slashchar{D}^2(y) +R \slashchar{D}(y)$ are also eigenstates 
of $\slashchar{D}(y)$ itself. 
Thus, we find the following eigenstates of $\slashchar{D}(y)$:
\bea
	&\ket[\psi_J] = (U_2\otimes U_N)
	\ket[1/2,+]\otimes \ket[J,J] \ ,\\
	&\ket[\psi_\bullet] = (U_2\otimes U_N) \ket[1/2,-]\otimes \ket[J,-J]. 
\ena
For the degenerate eigenstates of 
$\slashchar{D}^2(y) +R \slashchar{D}(y)$, 
we generally need to take a linear combination of them 
to find the eigenvectors of $\slashchar{D}(y)$. Thus, we consider  
\bea
	\ket[\psi_m] = (U_2\otimes U_N)
	\big(a_m \ket[1/2,+]\otimes \ket[J,m] +b_m \ket[1/2,-]\otimes \ket[J,m+1] \big)
	\ 
\ena
for $m=-J,-J+1,\cdots, J-1$.
By acting the Dirac operator on these states, we obtain
\bea
	\slashchar{D}\ket[\psi_m] 
	&= (U_2\otimes U_N) \bigg[ R\sigma^3  \otimes   L^3 -   \sigma^3  \otimes  |y|
	+ \frac{R}{2}(\sigma_+ \otimes L_-  +\sigma_-  \otimes   L_+ )\bigg]
	\nn
	&~~~\times \big(a_m \ket[1/2,+]\otimes \ket[J,m] +b_m \ket[1/2,-]\otimes \ket[J,m+1] \big)\ ,
\ena
where we utilized the properties of the unitary matrices $U_2$ and $U_N$
shown in appendix \ref{ApE}. 
The action of the $SU(2)$ generators on the right-hand side can also be
explicitly computed by using (\ref{def of states}).
Assuming that $\ket[\psi_m]$ are eigenstates  of $\slashchar{D}(y)$ 
with eigenvalues $\lambda_m$, 
we can obtain the following 
equations for $a_m$ and $b_m$:
\bea
	\lambda_m\vect[a_m,b_m] = \matt[Rm-|y|, R \sqrt{(J-m)(J+m+1)}, R \sqrt{(J-m)(J+m+1)},
	- R(m+1)  + |y| ] \vect[a_m,b_m]\ .	\label{eigeneq}
\ena
The characteristic equation reads
\bea
	0&=\det \matt[Rm-|y|-\lambda_m, 
R \sqrt{(J-m)(J+m+1)}, R \sqrt{(J-m)(J+m+1)},
	- R(m+1)  + |y| -\lambda_m ] \\
	&=\lambda_m^2 + R\lambda_m -|y|^2 +R(2m+1)|y|  -R^2J(J + 1)\ .
\ena
The eigenvalues are then given by
\bea
	2\lambda^{(\pm )}_m &= - R \pm \sqrt{R^2+4\{|y|^2 -R(2m+1)|y| +R^2J(J + 1)\}} \ .\label{EignVal}
\ena
The corresponding coefficients $a^{(\pm)}_m$ and $b^{(\pm)}_m$ 
have to satisfy (\ref{equation for am and bm}) 
to give a solution of (\ref{eigeneq}). 
The normalization condition for the state $\ket[\psi_m]$ also 
imposes (\ref{normaliz}).
Note that, without loss of generality, 
both $a^{(\pm)}_m$ and $b^{(\pm)}_m$ can be
set to be real numbers. 
Thus, the two equations (\ref{equation for am and bm}) and
(\ref{normaliz}) fully determine the states $\ket[\psi_m]$.

\section{Representation matrices of special unitary groups}
\subsection{Representation matrices of $SO(3)$ \label{ApE}}
In this appendix, we explicitly write down representation matrices 
of a $SO(3)$ rotation which transforms a general unit vector 
into the unit vector pointing the north pole.

Let $x$ be a general vector in $R^3$ parametrized as
\bea
	\left( \begin{array}{cc} x^1 \\ x^2 \\ x^3 \end{array} \right)
	= \left( \begin{array}{cc}  \sin \theta \cos\phi \\ \sin\theta\sin\phi \\ \cos\theta \end{array} \right)  ,
\label{unit vector on s2}
\ena
and $x_0$ be the unit vector pointing the north pole as 
\bea
	\left( \begin{array}{cc} x_0^1 \\ x_0^2 \\ x_0^3 \end{array} \right)
	= \left( \begin{array}{cc} 0 \\ 0 \\ 1 \end{array} \right)  .
\ena
We can consider an $SO(3)$ rotation which transforms $x$ to $x_0$, 
\begin{align}
x^i= (\Lambda^{-1})_{ij} x_0^j.
\end{align}
$\Lambda^{-1}$ is explicitly given by
\bea
	\Lambda^{-1} &=  
	 \left( \begin{array}{ccc} \cos\theta\cos\phi & -\sin\phi & \sin\theta\cos\phi \\  
 						\cos\theta\sin\phi & \cos\phi & \sin\theta\sin\phi \\ 
						-\sin\theta & 0 & \cos\theta \end{array} \right) 
	\left( \begin{array}{ccc} \cos\alpha & \sin\alpha & 0 \\  -\sin\alpha & \cos\alpha & 0 \\ 0 & 0 & 1 \end{array} \right)\\
	&=  
	\left( \begin{array}{ccc} \cos\phi & -\sin\phi & 0 \\  \sin\phi & \cos\phi & 0 \\ 0 & 0 & 1 \end{array} \right)
	\left( \begin{array}{ccc} \cos\theta & 0 & \sin\theta \\  0 & 1 & 0 \\ -\sin\theta & 0 & \cos\theta \end{array} \right)
	\left( \begin{array}{ccc} \cos\alpha & \sin\alpha & 0 \\  -\sin\alpha & \cos\alpha & 0 \\ 0 & 0 & 1 \end{array} \right)  .
\label{explicit form of Lambda}
\ena
Note that in defining $\Lambda^{-1}$ 
there is an ambiguity of $SO(2)$ rotations 
around the north pole.
This ambiguity is represented by the angle $\alpha$. 

Now, let us consider the action of this rotation on the generators of 
$SO(3)$ ($SU(2)$). 
Let $L^i$ $(i=1,2,3)$ be any irreducible representation matrices of 
$SU(2)$ generators. 
Since the representation matrices of the 
generators of Lie algebra are invariant tensors, 
there always exist unitary similarity transformations 
which undo the rotation of 
the vector index. Thus, there exists a unitary
matrix $U$ satisfying 
\begin{align}
{U}^\dagger L_i U=(\Lambda^{-1})_{ij} L_j.
\label{U undo}
\end{align}
If $\Lambda^{-1}$ is the 3-dimensional (vector) 
representation matrix of an element $g$ of $SU(2)$, 
the unitary matrix $U$
is given by the $N$-dimensional irreducible representation of 
the same element $g$, where $N$ is the dimension of 
the representation of $L^i$.

Below, we fix the ambiguity in the definition 
of $\Lambda^{-1}$ by putting $\alpha = \phi$.
From (\ref{explicit form of Lambda}), we find that 
the unitary matrix $U$ satisfying (\ref{U undo})
is given by
\bea
	U = e^{-i\phi L_3}e^{-i\theta L_2} e^{i\phi L_3}.
\label{Form of U}
\ena
This has another expression:
\bea
	U
	&=e^{zL^-}e^{-L^3\log(1+|z|^2)}e^{-\bar{z}L^+},
\label{form of U 2}
\ena
where $L^\pm = L^1 \pm iL^2$ and we introduced the stereographic coordinate
$(z, \bar{z})$ defined in (\ref{stereographic coordinate}).

\subsection{Representation matrices of $SO(5)$}
\label{Representation matrices of SO(5) rotations}
In this appendix, we show representation matrices of $SO(5)$ rotations. 

Let us first consider a unit vector in $\textbf{R}^5$, which can be parametrized 
in the polar coordinate as 
\bea
	\left( \begin{array}{cc} x^1 \\ x^2 \\ x^3 \\ x^4 \\ x^5 
\end{array} \right)
	= \left( \begin{array}{cc}  
\sin\theta\sin\phi\sin\psi\sin\chi \\ 
\sin\theta\sin\phi\sin\psi\cos\chi \\ 
\sin\theta\sin\phi\cos\psi         \\
\sin\theta\cos\phi\                \\ 
\cos\theta                     
 \end{array} \right)  .
\label{vector e}
\ena
We also consider the unit vector $x_0$ pointing the north pole given by
\bea
	\left( \begin{array}{cc} x_0^1 \\ x_0^2 \\ x_0^3 \\ x_0^4 \\ x_0^5
\end{array} \right)
	= \left( \begin{array}{cc} 0 \\ 0 \\ 0 \\ 0 \\ 1 \end{array} \right).
\ena
There exists $SO(5)$ rotation which transforms $x_0$ to $x$ as
\begin{align}
x^A = \Lambda^A{}_B x_0^B.
\label{e to e0 for fuzzy s4}
\end{align}
This transformation can be written as a product of some $SO(2)$ rotations.
Indeed, $\Lambda$ is given by a composition of 
a rotation on the 5-4 plane with angle $\theta$, 
a rotation on the 4-3 plane with angle $\phi$, 
a rotation on the 3-2 plane with angle $\psi$ and finally 
a rotation on the 2-1 plane with angle $\chi$.

We will write down the explicit form of $\Lambda$ in the following. 
We introduce the
generators of $SO(5)$ Lie algebra, $\Sigma_{A B},\ A, B \in\{1,2,3,4,5\}$,
which satisfies 
\beq
[\Sigma_{AB},\Sigma_{CD}]
=
\delta_{AD}\Sigma_{BC}+\delta_{BC}\Sigma_{AD}-\delta_{AC}\Sigma_{BD}
-\delta_{BD}\Sigma_{AC}\,.
\eeq
The fundamental (vector) and the spinor representation matrices  
of $\Sigma_{AB}$ are given by 
\begin{align}
&D_{\text{V}}(\Sigma_{AB})_{CD}
=
\delta_{AC}\delta_{BD}-\delta_{AD}\delta_{BC}\,, 
\nonumber\\
&D_{\text{S}}[\Sigma_{AB}]
=
\frac{1}{2}\Gamma_{AB}
=
\frac{1}{4}[\Gamma_A, \Gamma_B],
\end{align}
respectively. For example, 
in the vector representation, 
$\Sigma_{54}$ can be written as
\beq
D_{\text{V}}(\Sigma_{54})
=
\left(
\begin{array}{ccccc}
0&&&& \\
&0&&& \\
&&0&& \\
&&&0&-1 \\
&&&1&0
\end{array}
\right).
\eeq
This generates the rotation on the 5-4 plane with angle $\theta$, 
\beq
D_{\text{V}}(e^{-\theta\Sigma_{54}})
=
\left(
\begin{array}{ccccc}
1&&&& \\
&1&&& \\
&&1&& \\
&&&\cos\theta&\sin\theta \\
&&&-\sin\theta&\cos\theta
\end{array}
\right).
\eeq
Then, the rotation matrix $\Lambda$ in 
(\ref{e to e0 for fuzzy s4}) can be represented as 
\al{
\label{Jan9-3}
\Lambda
=
D_{\text{V}}(e^{-\chi\Sigma_{21}}e^{-\psi\Sigma_{32}}e^{-\phi\Sigma_{43}}e^{-\theta\Sigma_{54}})\,.
}
The spinor representation of the same group element,
\begin{align}
U
&=
D_{\text{S}}(e^{-\chi\Sigma_{21}}e^{-\psi\Sigma_{32}}e^{-\phi\Sigma_{43}}e^{-\theta\Sigma_{54}})\, \nonumber \\
&=
e^{-\chi\Gamma_{21}/2}e^{-\psi\Gamma_{32}/2}e^{-\phi\Gamma_{43}/2}
e^{-\theta\Gamma_{54}/2}\,
\label{U for fuzzy S^4}
\end{align}
satisfies the relation
\beq
\label{Jan9-1}
\Lambda_{AB}\Gamma_{B}=U^{\dagger}\Gamma_{A}U\,.
\eeq


\section{Hamiltonian method}
\subsection{Hamiltonian method for fuzzy $S^2$}
\label{appendix Hamiltonian method s2}
In this appendix, we compute the classical geometry of $S^2$ 
by using the Hamiltonian method.

The Hamiltonian for the fuzzy $S^2$ configuration (\ref{X for fuzzy s2})
is given by 
\begin{align}
H(y) = \frac{1}{2}(RL^i-y^i)^2 = \frac{1}{2}(R^2 J(J+1)+|y|^2) -R y^iL_i.
\end{align}
Thus, the problem is just reduced to diagonalizing the operator $y^iL_i$.
This is done in 
appendix~\ref{Spectrum of Dirac Operator for Fuzzy S2}, and the eigenstates 
are given by $U(y)| J, m \rangle$, where $U(y)$ is the $N$-dimensional 
representation matrix of the $SO(3)$ rotation defined in 
appendix~\ref{ApE}. The eigenvalues of $H(y)$ are given by
\begin{align}
\frac{1}{2}(R^2 J(J+1)+|y|^2) -R m |y|.
\end{align}
In particular, the ground state is given by $m=J$. In the large-$N$ limit, 
the ground state energy converges to 
\begin{align}
\frac{1}{2}(1-|y|)^2.
\end{align}
The classical geometry is defined as zeros of this function.
Thus, we find that the classical geometry is given by a unit sphere,
\begin{align}
{\cal M}= \{ y\in \textbf{R}^3 | |y|=1 \}.
\end{align}

The information metric and the Berry connection 
can also be computed in the similar 
way to the case of the Dirac operator. By using the differential of the 
ground state (\ref{dpsi}), one can quickly check that 
in the large-$N$ limit, the information metric
and Berry connection are equal to those obtained
in section \ref{Information metric and Berry connection for fuzzy S2}.

\subsection{Hamiltonian method for fuzzy $S^4$}
\label{Ham method for fuzzy s4}
The Hamiltonian for the matrices (\ref{Matrix for fuzzy S4}) is given by
\beq
	H(y)=\frac{1}{2}(1+|y|^2)-y_A X^A+\mathcal{O}(1/n),
\eeq
where $|y|=y_A y^A$. In order to find the spectrum of this Hamiltonian, we consider the specific $SO(5)$ rotation matrix $\Lambda$ that brings the vector in the direction of the pole $(0,0,0,0,|y|)$ to the position vector of a point $y\in\vb{R}^5$: $y_A\Lambda^A{}_B=|y|\delta_{B5}$. 
As shown in appendix~\ref{Representation matrices of SO(5) rotations}, 
for this rotation there exists a corresponding unitary operator $U$ which satisfies (\ref{Jan9-1}). It follows from the relation (\ref{Jan9-1}) that
\beq
	U^{\dagger\otimes n}(y_A X^A)U^{\otimes n}=|y|X_5.
\eeq
Using this relation we can diagonalize the Hamiltonian as
\beq
	H(y)
	=U^{\otimes n}\qty[\frac{1}{2}(1+|y|^2)-|y|X_5+\mathcal{O}(1/n)]U^{\dagger\otimes n}.
\eeq
Then we can easily find the ground states of $H(y)$ as
\beq
	\ket[0^{(\alpha)},y]=U^{\otimes n}\ket[{n}/{2},{n}/{2}-\alpha]\quad \alpha\in\qty{0,1,\ldots,n}.
\eeq
Here, the notation $| J,m \rangle $ introduced in 
section~\ref{Classical space for fuzzy S4} is used on the right-hand side. 
Note that $J=(n+1)/2$ in section~\ref{Classical space for fuzzy S4}, while 
$J=n/2$ in this appendix. This difference comes from the fact that the 
Dirac orator is defined in a bigger vector space. 
The eigenvalue of the ground states is
\beq
	E_0(y)=\frac{1}{2}(1-|y|^2)+\mathcal{O}(1/n).
\eeq
In the classical limit, the zeros of $E_0(y)$ are points such that $|y|=1$, and the classical space is indeed $S^4$ with unit radius.

Note that the structure of the ground state is common to that in the 
Dirac operator method. Hence, in the large-$N$ limit, the Berry connection
and the information metric for the Hamiltonian method are equivalent to 
those in the Dirac operator method.

\section{Derivation of useful relations for fuzzy $S^4$}
\label{Derivation of useful relations for fuzzy S4}
In this appendix, we prove some useful relations for fuzzy $S^4$.

We first prove (\ref{GG=1}). We first calculate as
\begin{align}
\sum_A (G^{(n)}_A)^2 
= 5n \1_{{\cal H}_n} + 2 {\cal O}.
\label{go}
\end{align}
Here, ${\cal O}$ is given by
\begin{align}
{\cal O}= \Gamma_A\otimes \Gamma_A \otimes \1_4 \otimes \1_4 + \cdots,
\label{operator o}
\end{align}
where $\cdots$ stands for all the symmetric 
permutations of the positions of $\Gamma_A$'s in the first term
(i.e. ${\cal O}$ has totally $n(n-1)/2$ terms). 
It is easy to see that ${\cal O}$ commutes with all of 
the $SO(5)$ generators, (\ref{symmetrized generator}). 
Thus, from Schur's lemma, ${\cal O}$ is proportional to the 
identity matrix on ${\cal H}_n$. The normalization constant 
can be fixed by acting ${\cal O}$ on the vector 
$|\eta_1 \rangle^{\otimes n}$. 
By using the representations (\ref{def of gamma}), we 
can easily prove that 
\begin{align} 
\sum_{a=1}^4 (\Gamma^a \otimes \Gamma^a) |\eta_1 \rangle 
\otimes |\eta_1 \rangle = 0.
\label{gamma gamma eta eta}
\end{align}
Then, we obtain 
\begin{align}
{\cal O}|\eta_1 \rangle^{\otimes n} = \frac{n(n-1)}{2}
|\eta_1 \rangle^{\otimes n}.
\end{align}
Hence, we find that 
\begin{align}
{\cal O} = \frac{n(n-1)}{2} \1_{{\cal H}_n}.
\label{o is 1}
\end{align}
Substituting this into (\ref{go}), we obtain
(\ref{GG=1}).

Next, we prove the following equations:
\begin{align}
\int d\Omega_4 (UP_+ U^{\dagger})^{\otimes n} 
&= \frac{16\pi^2 }{(n+2)(n+3)}\1_{{\cal H}_{n}},
\label{1st eq}
\\
\int d\Omega_4 x_A (UP_+ U^{\dagger})^{\otimes n} 
&= \frac{16\pi^2 }{(n+2)(n+3)(n+4)}G_A^{(n)}.
\label{2nd eq}
\end{align}
Here, the volume form $d\Omega_4$ shall be normalized as 
$\int d\Omega_4 = \frac{8\pi^2 }{3}$ and $x_A$ in the second
equation is defined in (\ref{vector e}). 
These equation follow from the fact that the integrations 
over $S^4$ produce only rotationally invariant tensors. 
Thus, from the structures of indices, we can see that 
the right-hand sides of (\ref{1st eq})
and (\ref{2nd eq}) are proportional to the identity matrix 
and $G_A^{(n)}$, respectively\footnote{Note that any contractions 
of the vector indices of Gamma matrices as in 
(\ref{operator o}) give the trivial identity matrix as shown in 
(\ref{o is 1}).}. 
Namely, we have 
\begin{align}
\int d\Omega_4 (UP_+ U^{\dagger})^{\otimes n} 
&= \alpha \1_{{\cal H}_{n}},
\label{1st eq 2}
\\
\int d\Omega_4 x_A (UP_+ U^{\dagger})^{\otimes n} 
&= \beta G_A^{(n)}.
\label{2nd eq 2}
\end{align}
The remaining task is to 
determine the proportionality constants $\alpha$ and $\beta$. 
$\alpha$ is determined by taking the trace of 
the both sides in (\ref{1st eq 2}).
Noting that 
\begin{align}
{\rm Tr}_{{\cal H}_{n}} 
(UP_+ U^{\dagger})^{\otimes n}
=
{\rm Tr}_{{\cal H}_{n}} 
P_+ ^{\otimes n}
=
{\rm Tr}_{{\cal H}^+_{n}} 
\1_{{\cal H}_n}
= {\rm dim} {\cal H}_n^+ =n+1,
\end{align}
we find that $\alpha $ is given as in (\ref{1st eq}).
$\beta$ is determined by multiplying $G_A^{(n)}$ and 
taking a summation over $A$ and finally taking the traces 
in the both sides of (\ref{2nd eq 2}). 
Because of (\ref{GG=1}), the right-hand side of 
(\ref{2nd eq 2}) becomes 
\begin{align}
\beta n(n+4){\rm Tr}_{{\cal H}_n} \1_{{\cal H}_n}
= \beta \frac{n (n+1)(n+2)(n+3)(n+4)}{6}.
\label{equating 1}
\end{align}
Because of (\ref{e to e0 for fuzzy s4}) and (\ref{Jan9-1}),
the left-hand side of (\ref{2nd eq 2}) becomes
\begin{align}
\int d\Omega_4 x^A \Lambda_{AB} {\rm Tr}_{{\cal H}_n} 
(G_B^{(n)}P_+^{\otimes n})
= 
\int d\Omega_4 {\rm Tr}_{{\cal H}_n} 
(G_5^{(n)}P_+^{\otimes n}) 
=
\int d\Omega_4 {\rm Tr}_{{\cal H}^+_n} 
(G_5^{(n)}) 
= \frac{8\pi^2}{3} n(n+1).
\label{equating 2}
\end{align}
By equating (\ref{equating 1}) and 
(\ref{equating 2}), we finally obtain (\ref{2nd eq}).

\section{Spin connections on $S^2$ and $S^4$}
\label{Spin connections on s2 and s4}
In this appendix, we list the spin connections on 
$S^2$ and $S^4$.
\subsection{Spin connections on $S^2$}
The standard round metric on $S^2$ in the stereographic coordinate 
is given by
\begin{align}
ds^2 = r^2 \frac{dzd\bar{z}}{(1+|z|^2)^2},
\end{align}
where $r$ is any positive constant corresponding to the radius of the sphere.
We introduce the vielbein by 
\begin{align}
e^+= \frac{r dz}{1+|z|^2}, \;\;\; 
e^-= \frac{r d\bar{z}}{1+|z|^2},
\end{align}
so that $ds^2 = e^+e^-$. The spin connection $\omega$ is determined by 
the equations $de^{\alpha} +\omega^{\alpha}{}_\beta \wedge e^\beta =0$.
In our case, the equations reduce to
\begin{align}
\omega^+{}_+ \wedge e^+ &=\frac{z}{r}e^- \wedge e^+, \nonumber\\
\omega^-{}_- \wedge e^- &=\frac{\bar{z}}{r}e^+ \wedge e^-.
\end{align}
The solution to these equations is given by
\begin{align}
\omega^+{}_+ =
-\omega^-{}_- = \frac{1}{r}(ze^--\bar{z}e^+) = \frac{zd\bar{z}-\bar{z}dz}
{1+|z|^2}.
\end{align}

\subsection{Spin connections on $S^4$}
The standard round metric on $S^4$ in the polar coordinate 
is given by
\begin{align}
ds^2
=r^2
\qty(d\theta^2+\sin^2\theta\,d\phi^2+\sin^2\theta\sin^2\phi\,d\psi^2
+\sin^2\theta\sin^2\phi\sin^2\psi\,d\chi^2),
\end{align}
where $r$ is the radius of $S^4$. We define the vielbein by 
\begin{align}
&e^1 = r \sin\theta \sin\phi \sin\psi d\chi, \nonumber\\
&e^2 = r \sin\theta \sin\phi d\psi, \nonumber\\
&e^3 = r \sin\theta d\phi, \nonumber\\
&e^4 = r d\theta.
\label{vielbein on s4}
\end{align}
By solving the equations 
$de^\alpha + \omega^\alpha{}_\beta \wedge e^\beta =0$, 
we obtain the spin connection as 
\begin{align}
&\omega_{12}= \cos\psi d\chi, \nonumber\\
&\omega_{13}= \cos\phi\sin\psi d\chi, \nonumber\\
&\omega_{14}= \cos\theta\sin\phi\sin\psi d\chi, \nonumber\\
&\omega_{23}= \cos\phi d\psi, \nonumber\\
&\omega_{24}= \cos\theta\sin \phi d\psi, \nonumber\\
&\omega_{34}= \cos\theta d\phi.
\label{omega for s4}
\end{align}

\section{Quantization maps for Laplacians on $S^2$ and $S^4$}
In this appendix, we derive the mapping rules 
(\ref{mapping delta s2}) and (\ref{mapping delta s4})
for Laplace operators on $S^2$ and $S^4$, respectively.
\subsection{Laplacian on $S^2$}
\label{Laplacians on S^2}
Here, we derive (\ref{mapping delta s2}). 
From the mapping rule (\ref{fij for s2}), we have
\begin{align}
(\hat{\Delta f})_{ij}
=({\cal D}_a^2 \psi_j, f\psi_i)+
( \psi_j, f {\cal D}_a^2\psi_i)+
2({\cal D}_a \psi_j, f{\cal D}_a\psi_i),
\label{delta f partial integrated}
\end{align}
where we have used partial integrations.
The first and the second terms in 
(\ref{delta f partial integrated}) can be evaluated 
with the formula
\begin{align}
{\cal D}_a^2\psi_i = 
-\frac{J}{r^2} \psi_i.
\label{ddpsi=psi}
\end{align}
This is obtained as follows. Since 
${\cal D}\!\!\!\!/ \; \psi_i =0$, we have
\begin{align}
{\cal D}_a^2\psi_i = 
(\sigma^{a}\sigma^{b}-\sigma^{ab}) {\cal D}_a {\cal D}_b \psi_i
=-\frac{1}{2}\sigma^{ab} [{\cal D}_a, {\cal D}_b] \psi_i
=-\frac{1}{2}\sigma^{ab}\left( 
\frac{1}{4}R_{abcd}\sigma^{cd}-iF_{ab} \right)\psi_i.
\label{ddpsi}
\end{align}
For $S^2$ with radius $r$, the curvature tensor is given by 
\begin{align}
R_{abcd}= \frac{1}{r^2}(\delta_{ac}\delta_{bd}-\delta_{ad}\delta_{bc}),
\label{curvature on S2}
\end{align}
and $F_{ab}=e_a^{\mu}e_b^{\nu}F_{\mu \nu}$ is obtained from
(\ref{F for fuzzy S2}) as
\begin{align}
F_{12}=\frac{N}{2r^2}.
\label{Fab on S2}
\end{align}
Substituting (\ref{curvature on S2}) and (\ref{Fab on S2}) into 
(\ref{ddpsi}), we obtain (\ref{ddpsi=psi}). The third term in
(\ref{delta f partial integrated}) is evaluated by using
\begin{align}
{\cal D}_{\pm}\psi_i = \frac{1}{r} \Lambda_{\mp k} (L_k)_{ij} \psi_j.
\end{align}
These equations follow from (\ref{nabla psi}) and (\ref{U undo}).
By using the relation
\begin{align}
\sum_{a=1}^2 \Lambda_{ak}\Lambda_{ak'} = \delta_{kk'}-x^k x^{k'},
\end{align}
where $x^k$ is defined in (\ref{unit vector on s2}), 
we find that the third term in 
(\ref{delta f partial integrated}) is given by 
\begin{align}
-\frac{2J^2}{r^2}\hat{f}_{ij} + \frac{2}{r^2}(L_k\hat{f}L_k)_{ij}.
\end{align}
From this and (\ref{ddpsi}), we obtain
(\ref{mapping delta s2}).

\subsection{Laplacian on $S^4$}
\label{Laplacians on S^4}
The mapping rule for the Laplace operator on $S^4$ can be obtained 
in a similar way as the case of $S^2$. 
First, it is easy to see that 
\begin{align}
\frac{1}{2}\Gamma^{ab}F_{ab}^k 
= \frac{4i}{r^2}D_S(J^k).
\label{GammaFS4}
\end{align}
The curvature tensor of $S^4$ with radius $r$ is given by the 
same form as (\ref{curvature on S2}), where the 
indices $a,b,c,d$ run from $1$ to $4$ for $S^4$. 
Then, from (\ref{GammaFS4}), the relation
\begin{align}
\Gamma^{ab}[D_a, D_b]\psi_i^{Jm}=
\frac{4J-2}{r^2}\psi_i^{Jm}
\label{Laps41}
\end{align}
holds. From (\ref{zero eigen spinors on s4}), it is also easy to obtain
\begin{align}
D_a \psi_i^{Jm} = \frac{\sqrt{N}}{2r(n+1)}
\sum_{s,\gamma}
C^{Jm}_{\frac{1}{2}s J-\frac{1}{2}\gamma}
|1/2, s \rangle \langle i | U^{\otimes n}
G_a^{(n)}|J-1/2,\gamma \rangle.
\label{Dpsis4}
\end{align}
By using (\ref{Laps41}) and (\ref{Dpsis4}), we can evaluate
the Toeplitz operator for the Laplacian on $S^4$ defined by 
\begin{align}
\hat{\Delta f}_{ij} 
&= \frac{1}{n+2}\sum_m \frac{3(n+1)^2}{8\pi^2}
\int d\Omega_4 (\psi_j^{Jm})^\dagger \psi_i^{Jm} \Delta f.
\end{align}
By integrating by parts, this is given by the sum of terms 
such as $\int d\Omega_4 (\psi_j^{Jm})^\dagger (D_a^2 \psi_i^{Jm}) f$,
$\int d\Omega_4 (D_a^2\psi_j^{Jm})^\dagger \psi_i^{Jm} f$
and $\int d\Omega_4 (D_a \psi_j^{Jm})^\dagger (D_a \psi_i^{Jm}) f$.
The first two can be evaluated by noting that
\begin{align}
D_aD_a \psi_i^{Jm} = (\Gamma^a \Gamma^b -\Gamma^{ab})D_aD_b \psi_i^{Jm}
= -\frac{1}{2}\Gamma^{ab}[D_a,D_b]\psi_i^{Jm},
\end{align}
and using (\ref{Laps41}), while the third term can be calculated 
with (\ref{Dpsis4}). These calculations lead to 
the mapping rule (\ref{mapping delta s4}).


\begin{thebibliography}{99}
\bibitem{Banks:1996vh} 
  T.~Banks, W.~Fischler, S.~H.~Shenker and L.~Susskind,
  Phys.\ Rev.\ D {\bf 55}, 5112 (1997).


\bibitem{Ishibashi:1996xs} 
  N.~Ishibashi, H.~Kawai, Y.~Kitazawa and A.~Tsuchiya,
  Nucl.\ Phys.\ B {\bf 498}, 467 (1997).

\bibitem{deWit:1988wri} 
  B.~de Wit, J.~Hoppe and H.~Nicolai,
  Nucl.\ Phys.\ B {\bf 305}, 545 (1988).

\bibitem{Arnlind:2010ac} 
  J.~Arnlind, J.~Hoppe and G.~Huisken,
  J.\ Diff.\ Geom.\  {\bf 91}, no. 1, 1 (2012).


\bibitem{Madore:1991bw} 
  J.~Madore,
  Class.\ Quant.\ Grav.\  {\bf 9}, 69 (1992).

\bibitem{Balachandran:2001dd} 
  A.~P.~Balachandran, B.~P.~Dolan, J.~H.~Lee, X.~Martin and D.~O'Connor,
  J.\ Geom.\ Phys.\  {\bf 43}, 184 (2002).

\bibitem{Connes:1997cr} 
  A.~Connes, M.~R.~Douglas and A.~S.~Schwarz,
  JHEP {\bf 9802}, 003 (1998).

\bibitem{Arnlind:2006ux} 
  J.~Arnlind, M.~Bordemann, L.~Hofer, J.~Hoppe and H.~Shimada,
  JHEP {\bf 0906}, 047 (2009).


\bibitem{Castelino:1997rv} 
  J.~Castelino, S.~Lee and W.~Taylor,
  Nucl.\ Phys.\ B {\bf 526}, 334 (1998).

\bibitem{Sperling:2017dts} 
  M.~Sperling and H.~C.~Steinacker,
  J.\ Phys.\ A {\bf 50}, no. 37, 375202 (2017).


\bibitem{Bordemann:1993zv} 
  M.~Bordemann, E.~Meinrenken and M.~Schlichenmaier,
  Commun.\ Math.\ Phys.\  {\bf 165}, 281 (1994).

\bibitem{Ma-Marinescu}
X.~Ma and G.~Marinescu, 
J.\ Geom.\ Anal.\ {\bf 18}, no. 2, 565-611 (2008).

\bibitem{Hasebe:2010vp} 
  K.~Hasebe,
  SIGMA {\bf 6}, 071 (2010).

\bibitem{Hasebe:2015htd} 
  K.~Hasebe,
  Int.\ J.\ Mod.\ Phys.\ A {\bf 31}, no. 20n21, 1650117 (2016).

\bibitem{Hasebe:2017myo} 
  K.~Hasebe,
  arXiv:1712.07767 [hep-th].



\bibitem{Ishiki:2015saa} 
  G.~Ishiki,
  Phys.\ Rev.\ D {\bf 92}, no. 4, 046009 (2015).

\bibitem{Schneiderbauer:2016wub} 
  L.~Schneiderbauer and H.~C.~Steinacker,
  J.\ Phys.\ A {\bf 49}, no. 28, 285301 (2016).

\bibitem{Berenstein:2012ts} 
  D.~Berenstein and E.~Dzienkowski,
  Phys.\ Rev.\ D {\bf 86}, 086001 (2012).

\bibitem{deBadyn:2015sca} 
  M.~H.~de Badyn, J.~L.~Karczmarek, P.~Sabella-Garnier and K.~H.~C.~Yeh,
  JHEP {\bf 1511}, 089 (2015).

\bibitem{Karczmarek:2015gda} 
  J.~L.~Karczmarek and K.~H.~C.~Yeh,
  JHEP {\bf 1511}, 146 (2015).

\bibitem{Ishiki:2016yjp} 
  G.~Ishiki, T.~Matsumoto and H.~Muraki,
  JHEP {\bf 1608}, 042 (2016).

\bibitem{TachyonA} 
  T.~Asakawa, G.~Ishiki, T.~Matsumoto, S.~Matsuura and H.~Muraki,
  arXiv:1804.00161[hep-th].

\bibitem{TachyonB} 
  S.~Terashima,
  arXiv:1804.00647[hep-th].

\bibitem{Terashima:2005ic} 
  S.~Terashima,
  JHEP {\bf 0510}, 043 (2005).

\bibitem{Asakawa:2001vm} 
  T.~Asakawa, S.~Sugimoto and S.~Terashima,
  JHEP {\bf 0203}, 034 (2002).

\bibitem{Hashimoto:2005qh} 
  K.~Hashimoto and S.~Terashima,
  JHEP {\bf 0602}, 018 (2006).

\bibitem{Seiberg:1999vs} 
  N.~Seiberg and E.~Witten,
  JHEP {\bf 9909}, 032 (1999).

\bibitem{Guralnik:2000pb} 
  Z.~Guralnik and S.~Ramgoolam,
  JHEP {\bf 0102}, 032 (2001).

\bibitem{Kaneko:2017zeo} 
  Y.~Kaneko, H.~Muraki and S.~Watamura,
  Class.\ Quant.\ Grav.\  {\bf 35}, no. 5, 055009 (2018).


\bibitem{Steinacker:2016vgf} 
  H.~C.~Steinacker,
  JHEP {\bf 1612}, 156 (2016).

\bibitem{Hanada:2005vr} 
  M.~Hanada, H.~Kawai and Y.~Kimura,
  Prog.\ Theor.\ Phys.\  {\bf 114}, 1295 (2006).

\end{thebibliography}
\end{document}